\documentclass[apj,twocolumn,twocolappendix,floatfix]{openjournal}

\DeclareRobustCommand{\VAN}[3]{#2}
\let\VANthebibliography\thebibliography
\def\thebibliography{\DeclareRobustCommand{\VAN}[3]{##3}\VANthebibliography}

\usepackage{amsmath}	% Advanced maths commands
\usepackage{amssymb}	% Extra maths symbols
\usepackage{balance}
\usepackage{booktabs}
\usepackage{lipsum}
\usepackage[T1]{fontenc}
\usepackage{graphicx}	% Including figure files
\usepackage{natbib}
\usepackage[breaklinks,colorlinks,citecolor=blue,urlcolor=blue,linkcolor=blue]{hyperref}
\usepackage{cleveref}
\usepackage{nicematrix}
\usepackage{nicematrix}
\usepackage{multirow}
\usepackage{natbib}
\usepackage{newtxtext,newtxmath}
\usepackage{orcidlink}
\usepackage{pifont}% http://ctan.org/pkg/pifont
\usepackage[nobottomtitles]{titlesec}
\usepackage{natbib}
\usepackage{bibentry}

\usepackage{xparse}
\usepackage{ifthen}
\usepackage{gensymb}
\usepackage{ctable}
\titleformat{\section}{\filcenter\MakeUppercase}{\thesection.}{0.5em}{}
\usepackage{hyperref}
\usepackage{savesym}
\savesymbol{tablenum}
\usepackage{siunitx}
\restoresymbol{SIX}{tablenum}
\usepackage{needspace}
\usepackage{overpic}
\usepackage{xcolor}
\usepackage{lipsum}
\usepackage{pifont}
\usepackage{siunitx}
\usepackage[flushleft]{threeparttable}

\newcommand{\Gaia}{\textit{Gaia}}
\newcommand{\TESS}{\textit{TESS}}

\newcommand{\cmark}{\ding{51}}%
\newcommand{\xmark}{\ding{55}}%

\begin{document}
% Title of the paper, and the short title which is used in the headers.
% Keep the title short and informative.
\title[Astronomical Cardiology II]{Astronomical Cardiology II: A Search For Heartbeat Stars Using APOGEE and \textit{TESS}}

% The list of authors, and the short list which is used in the headers.
% If you need two or more lines of authors, add an extra line using \newauthor

\author{\vspace{-1.3cm}Jowen Callahan$^{1}$,
D. M. Rowan\,\orcidlink{0000-0003-2431-981X}$^{2\dagger}$,
C. S. Kochanek\ $^{1,3}$,
M. M. Fausnaugh\,\orcidlink{0000-0002-9113-7162}$^{4}$
}

\affiliation{$^{1}$Department of Astronomy, The Ohio State University, 140 West 18th Avenue, Columbus, OH, 43210, USA\\
$^{2}$Department of Astronomy, University of California Berkeley, Berkeley CA 94720, USA\\
$^{3}$Center for Cosmology and Astroparticle Physics, The Ohio State University, 191 W. Woodruff Avenue, Columbus, OH, 43210, USA\\
$^{4}$Department of Physics \& Astronomy, Texas Tech University, Lubbock, TX 79409-1051, USA \\
$^{\dagger}$NHFP Hubble Fellow
}

% Abstract of the paper
\begin{abstract}
Stellar binaries in short-period, highly eccentric systems with significant tidal deformations near pericenter, also known as heartbeat stars, are a laboratory for studying dynamical tides and oscillations in stars. We identify 50 heartbeat stars using \textit{TESS} light curves of stars identified as binaries using SDSS APOGEE. We fit their phase-folded \textit{TESS} light curves with an analytic model to measure their orbital periods, eccentricities, inclinations, and arguments of periastron. We measure the mass function of targets with enough APOGEE radial velocity observations to obtain a constraint on the secondary mass. We confirm our previous results that the non-giant heartbeat stars have started to evolve off the main sequence and that the fraction of (near) main sequence binaries that are heartbeat stars rises rapidly with effective temperature.
\end{abstract}

\maketitle

\section{Introduction}

Heartbeat stars (HBs) are a subclass of detached binary stars with short periods ($P\lesssim100$ days) and high eccentricities ($e\gtrsim0.2$) that earn their name from the structure of their light curves. At periastron, the stars pass close enough to be tidally deformed, leading to the fluctuations in brightness that give the light curves their namesake electrocardiogram-like shape \citep{Thompson2012}. 

Using \textit{Kepler} \citep{2010Borucki}, \citet{Welsh2011} detected the first HB, KOI-54, which also shows tidally excited oscillations (TEOs) between periastron passages \citep{Fuller2012, Burkart2012, Leary2014}. Over the course of the \textit{Kepler} mission, hundreds of main sequence (MS) HBs were discovered \citep{Thompson2012, Hambleton2013, Kirk2016, Shporer2016}. The \textit{Kepler} HBs are predominantly MS binaries with low variability amplitudes ($\lesssim 0.1$~mmag), but evolved HBs on longer-period orbits with larger-amplitude variability have also been discovered with the Optical Gravitational Lensing Experiment \citep[OGLE, ][]{1992Udalski} in the Magellanic Clouds \citep{Wrona2022a, Wrona2022b}.

The \textit{Transiting Exoplanet Survey Satellite} \citep[\textit{TESS}, ][]{Ricker2015} led to a second boom in HB discoveries. Early detections occurred as a byproduct of planet searches \citep{Wheel2019}, investigations of known variable stars \citep{THARINDU2019, THARINDU2021, Merc2021, Paunzen2021, Barb2022, SK2022, Wang2023}, and searches for new \textit{TESS} variables \citep{Murphy2020, Kochukhov2021, Sharma2022, PMR2024}. Later, systematic searches for HBs with \textit{TESS} began to return hundreds of new detections \citep{SK2021, Li2024a, Li2024b, Li2025, Sol2025, 2025Zhou}. 
\defcitealias{Me}{C25}

In our previous work \citep[][hereafter \citetalias{Me}]{Me}, we identified 112 HBs starting from the \textit{Gaia} spectroscopic binary catalogs \citep{GAIA2023SB} with \textit{TESS} photometry. We used the \citet{Kumar1995} analytic model to fit the light curves and measure periods, eccentricities, inclinations, and the arguments of periastron. While the HB solutions showed that many of the \textit{Gaia} orbital parameters were inaccurate for short-period, eccentric binaries, the \textit{Gaia} non-single stars (NSS) catalog provided an effective starting sample for identifying rare subclasses of binaries. For two systems with \textit{Gaia} double-lined spectroscopic binary orbital solutions, we jointly fit the \textit{TESS} light curve and NSS solution to measure the masses and radii of both components.

When we examined the resulting HB population, we found two significant trends for the MS HBs. First, the HBs are shifted far enough from the MS that most primaries must have begun to evolve towards becoming subgiants. Second, the abundance of MS HBs is highest on the upper MS and drops rapidly at lower temperatures. HB amplitudes reflect a competition between the rate of stellar evolution and the amount of damping \citep{2025MacLeod}. Early-type MS stars evolve quickly, which drives larger variability amplitudes. Their radiative envelopes also produce less damping than the convective envelopes of cooler late-type stars below the Kraft break \citep{1967Kraft, 1962Schatzman, 2024Beyer}. Together, these effects lead to a decreasing frequency of HBs at lower temperatures in any ``amplitude-limited" search, like what was done in \citetalias{Me}.

Here we carry out an analysis similar to \citetalias{Me} using the Sloan Digital Sky Survey's Apache Point Observatory Galactic Evolution Experiment \citep[SDSS APOGEE, ][]{2000York, 2011Eisenstein, 2017Blanton, 2017Majewski, 2022AA, 2025SDSS} and \textit{TESS} to identify HB binaries and to test the MS trends we found in our previous work. We discuss the selection process and the \textit{TESS} light curves in \S\ref{sec:HBSS}. In \S\ref{sec:RV}, we use the APOGEE radial velocity (RV) observations of our HBs to measure the binary mass functions and to estimate the secondary masses where possible. We discuss the results and our conclusions in \S\ref{sec:D}.

\section{Heartbeat Star Search}
\label{sec:HBSS}

APOGEE obtains its data in a series of visits that provide a very sparse set of RV measurements. \citet{2018Badenes} showed that systems with significant velocity scatter in these measurements must be binaries, so we did our selection based on their criterion of ${\tt vscatter}\ge3$ km/s. We were particularly interested in improving the statistics for MS HBs, so we explicitly removed giants by selecting systems with surface gravity $\log g\ge3.25$. This resulted in a sample of 38,253 MS binaries. We restrict our analysis to the 31,548 systems with \TESS{} magnitude $T< 13.5$, for which we use light curves from the Quick-Look Pipeline \citep[QLP, ][]{Huang2020} covering sectors 1--79. Light curves for the 6,705 fainter systems in our sample could be extracted directly from the full-frame images using difference-imaging pipelines \citep[e.g.,][]{2021Fausnaugh, 2023Fausnaugh}, but the reduced photometric precision at these magnitudes precludes the detailed modeling presented here. On average, there are six sectors of data available, leaving 194,532 unique light curves. As in \citetalias{Me}, we model each sector independently when searching for HBs.

We use the same semi-automated search method described in \citetalias{Me}. To summarize, we start by searching for periodicity with a Lomb-Scargle periodogram \citep[LS, ][]{Lomb1976, Scargle1982}. We refine this period by applying phase-dispersion-minimization \citep[PDM, ][]{Stellingwerf1978} around $P_{LS}$, the period of maximum LS power. Since the LS periodogram can return a harmonic of the period, we consider candidate orbital periods of $P_{LS}$, $2P_{LS}$, and $3P_{LS}$. Finally, before fitting our light curves using {\tt SciPy} {\tt curve\_fit} \citep{2020SciPy-NMeth}, we bin the phase-folded light curves into half-hour time bins to reduce the computational effort. 

We fit the phase-folded light curves using the \citet{Kumar1995} analytic model \citep{Wrona2022b} and a Trust Region Reflective algorithm. The flux is modeled as

\begin{equation}
    F=Z+S\frac{1-3\sin^2{i}\sin^2{(\nu+\omega)}}{(1-e\cos{E})^3},
	\label{eq:Flux}
\end{equation}where $S$ sets the amplitude, $Z$ is the mean flux, $\omega$ is the argument of periastron, $i$ is the inclination, and $e$ is the orbital eccentricity. The true anomaly is

\begin{equation}
    \nu=2\tan^{-1}\biggl({\frac{\sqrt{1+e}}{\sqrt{1-e}}{\tan{\frac{E}{2}}}}\biggr),
	\label{eq:True}
\end{equation}and the eccentric anomaly

\begin{equation}
    E-e\sin{E} = \frac{2\pi(t-t_{0})}{P},
	\label{eq:Ecc}
\end{equation}is determined by the period, $P$, and the epoch of periastron, $t_0$. Additionally, we repeat the optimization twice with different starting points to avoid falling into local minima, so for 31,548 binaries we tried 1,167,192 models in total.

\begin{figure}
	\includegraphics[width=\columnwidth]{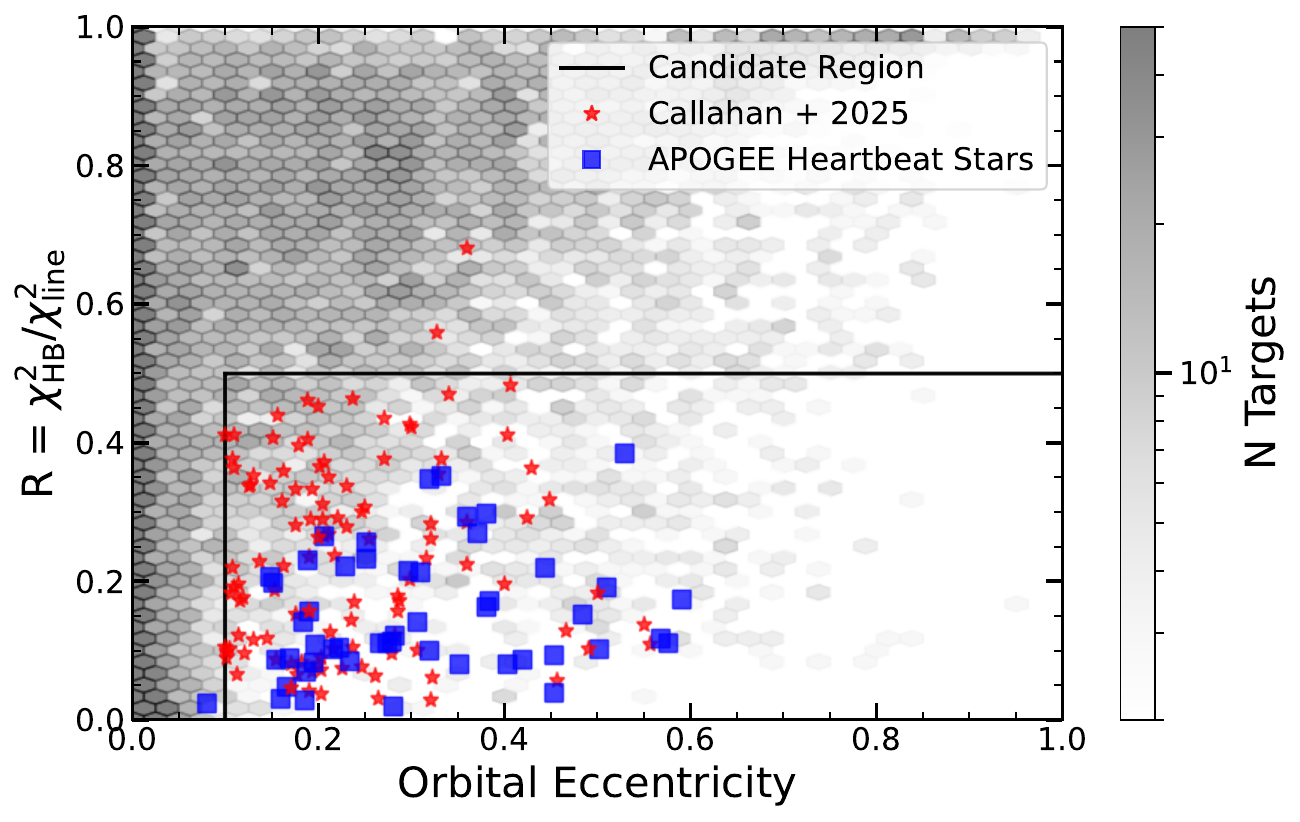}
    \caption{The density distribution of $R=\chi_{HB}^2/\chi_{line}^2$ and eccentricity for the fits to the APOGEE stars (grey background). We used a candidate region based on the HBs identified using \textit{Gaia} spectroscopic binaries identified in \citet{Me} (stars). The final models of the APOGEE HBs (squares) can lie outside the candidate region when masking eclipses leads to changes in the model.}
    \label{fig:Final}
\end{figure}

\begin{figure*}
    \centering
    \begin{tabular}{ccc}
        \includegraphics[width=0.33\textwidth]{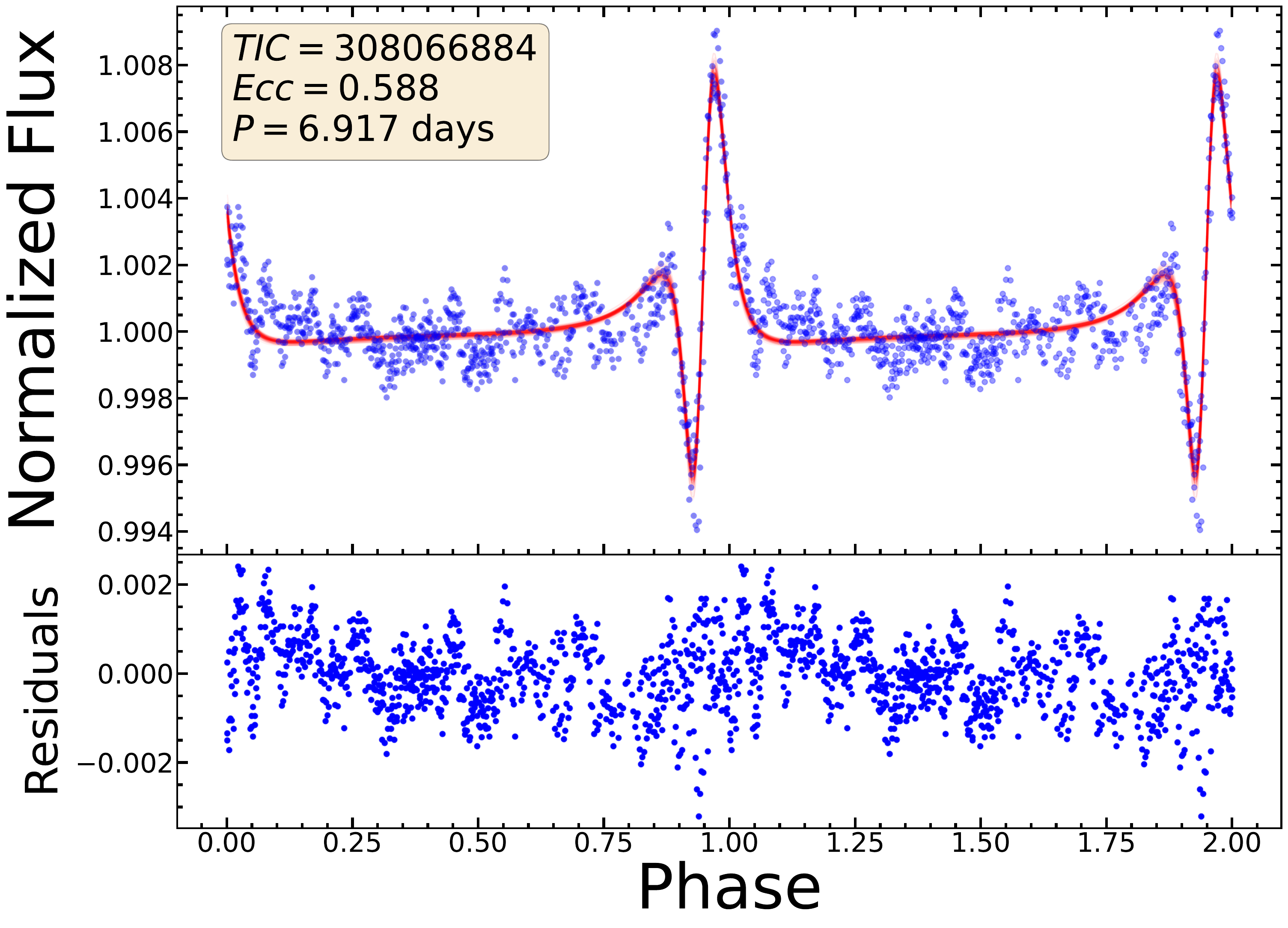} &       
       \includegraphics[width=0.33\textwidth]{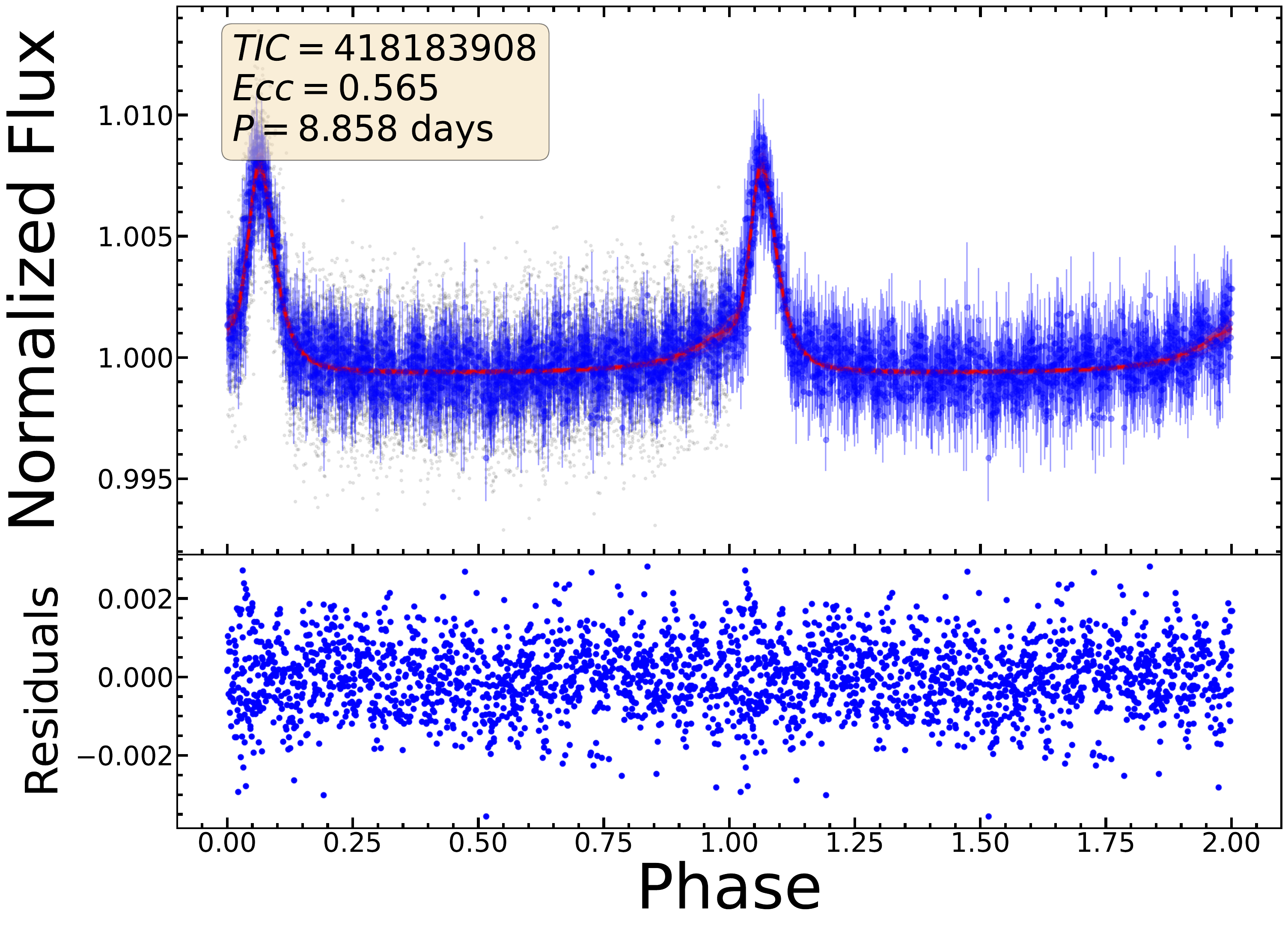} &
        \includegraphics[width=0.33\textwidth]{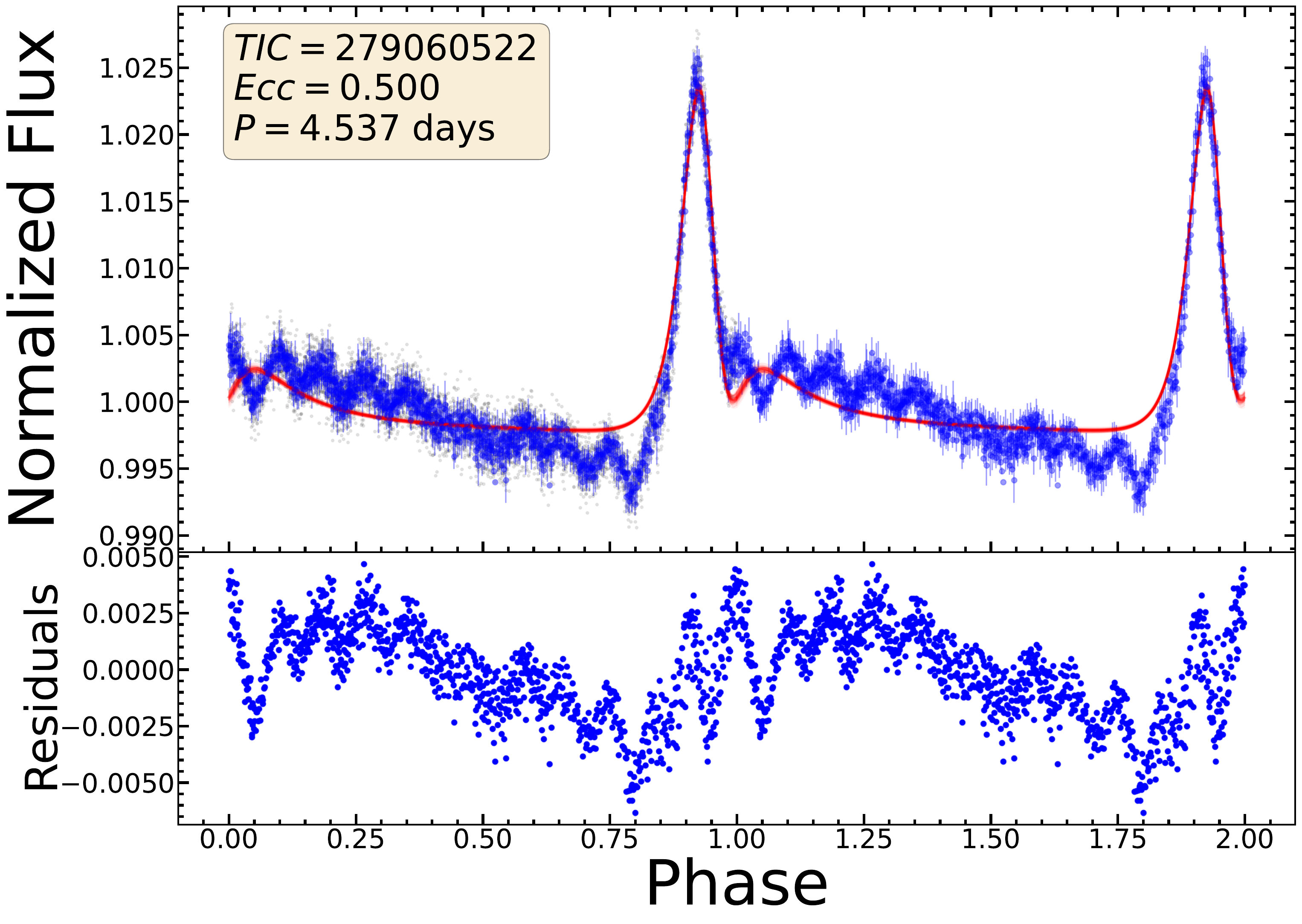} \\
       
       \includegraphics[width=0.33\textwidth]{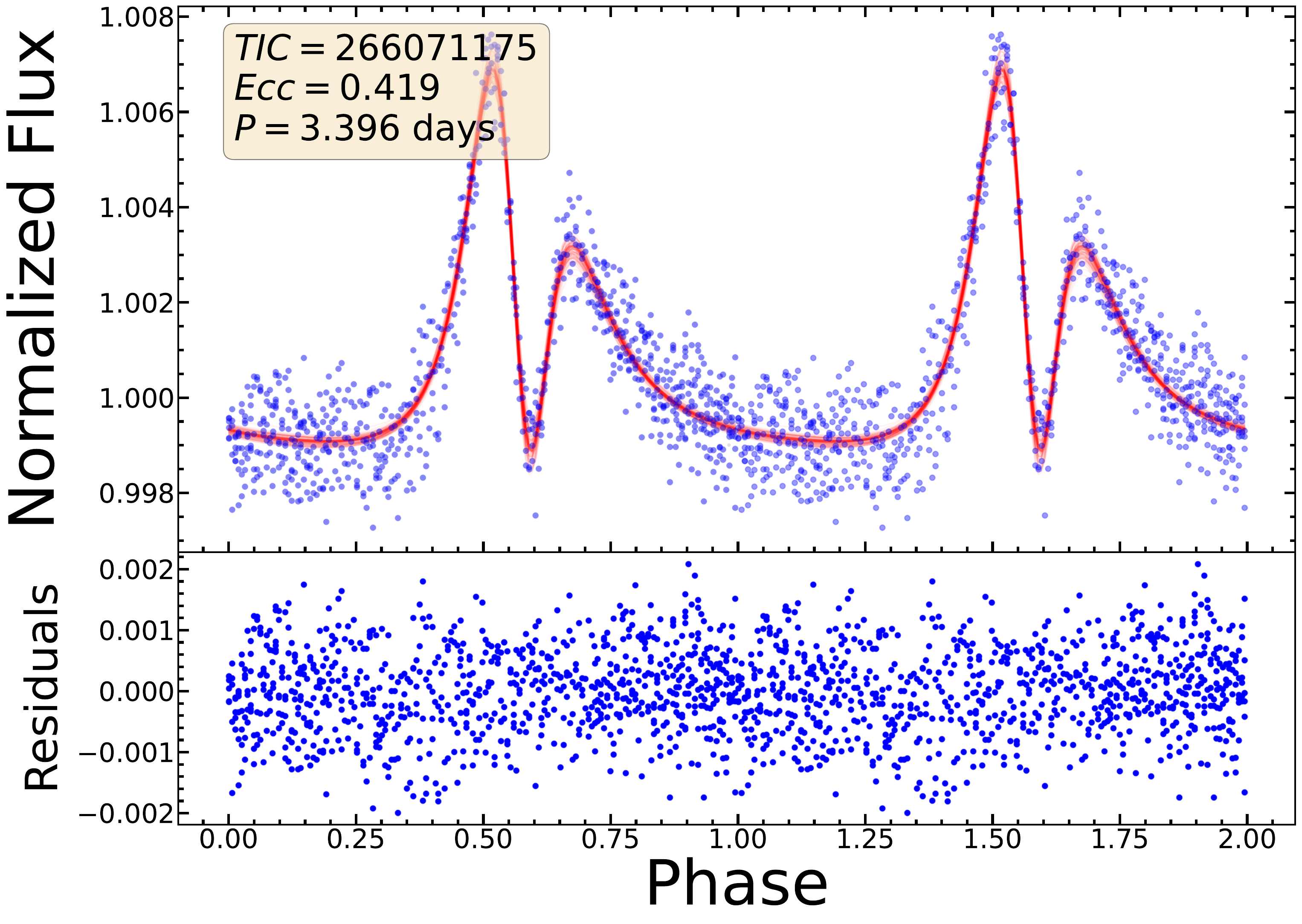} &
	\includegraphics[width=0.33\textwidth]{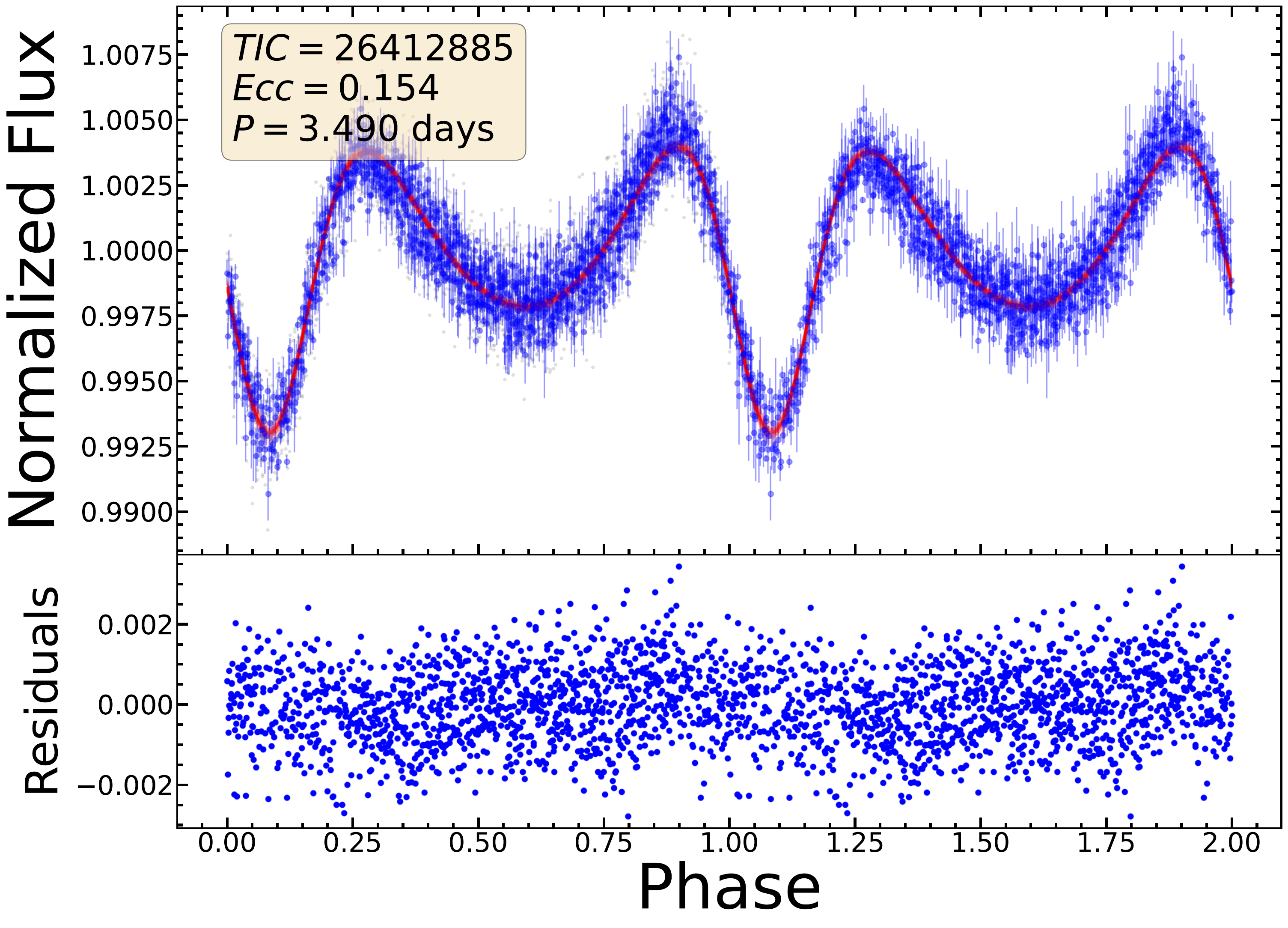} &       
       \includegraphics[width=0.33\textwidth]{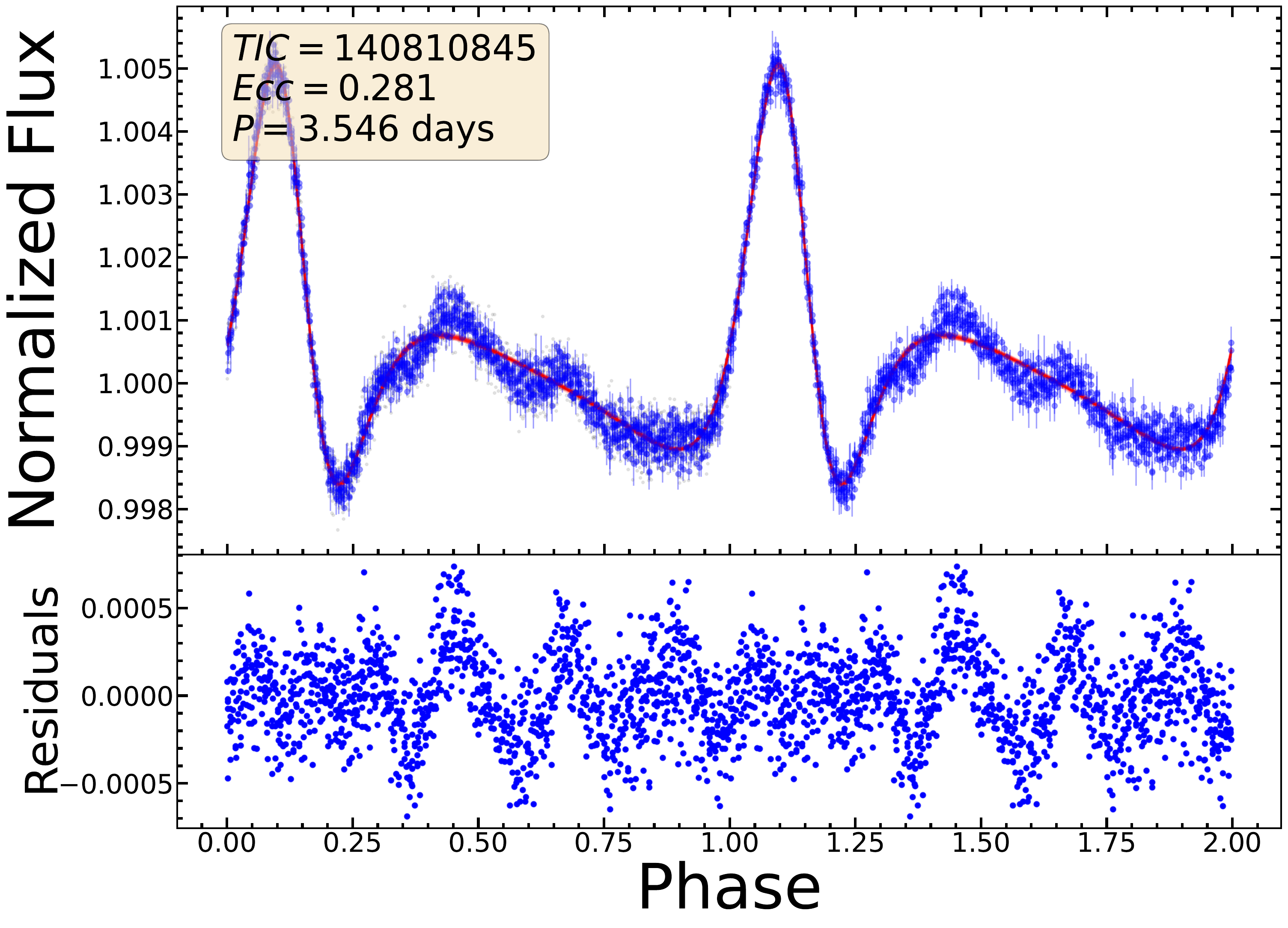} \\
      
        \includegraphics[width=0.33\textwidth]{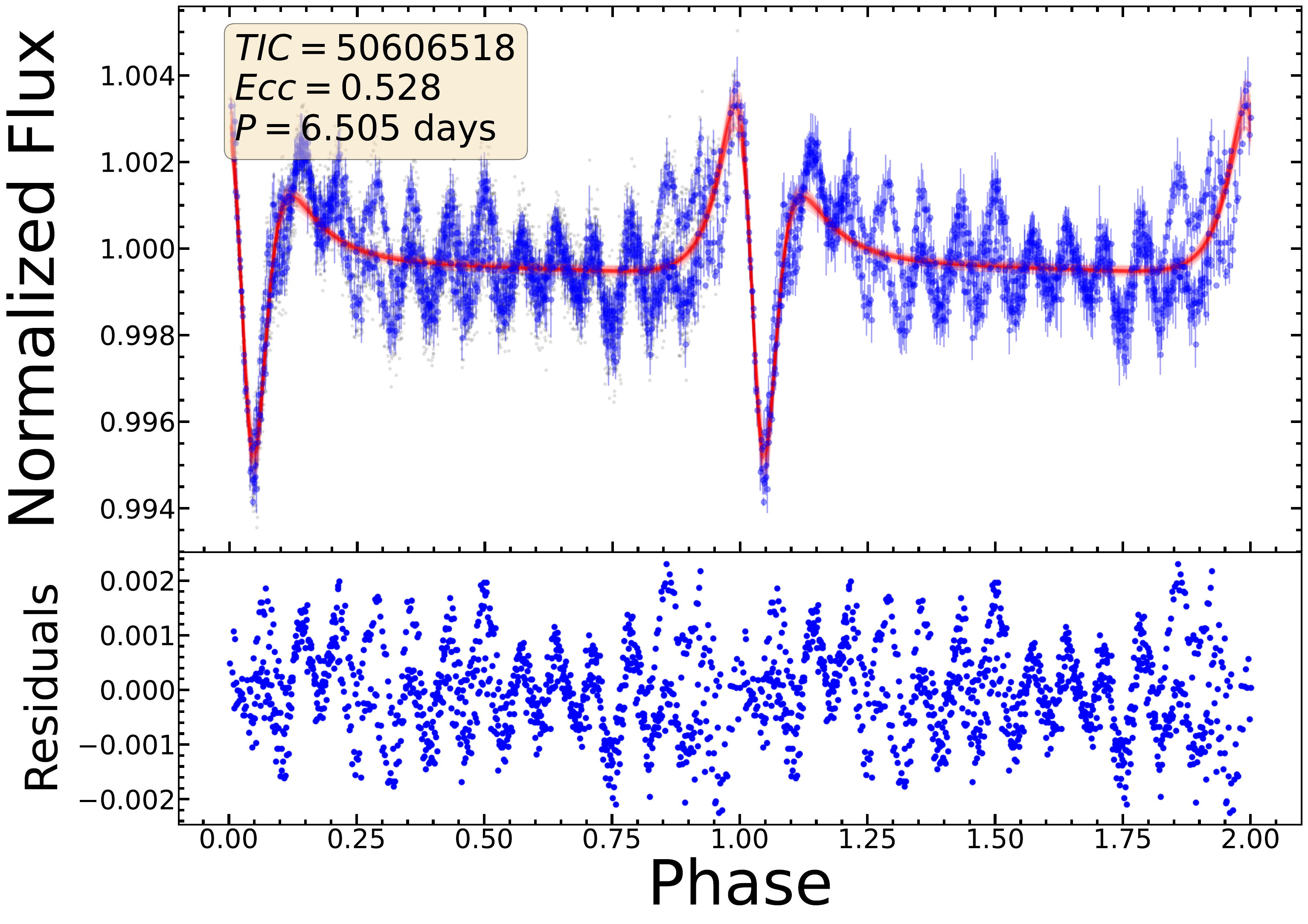} &
       	\includegraphics[width=0.33\textwidth]{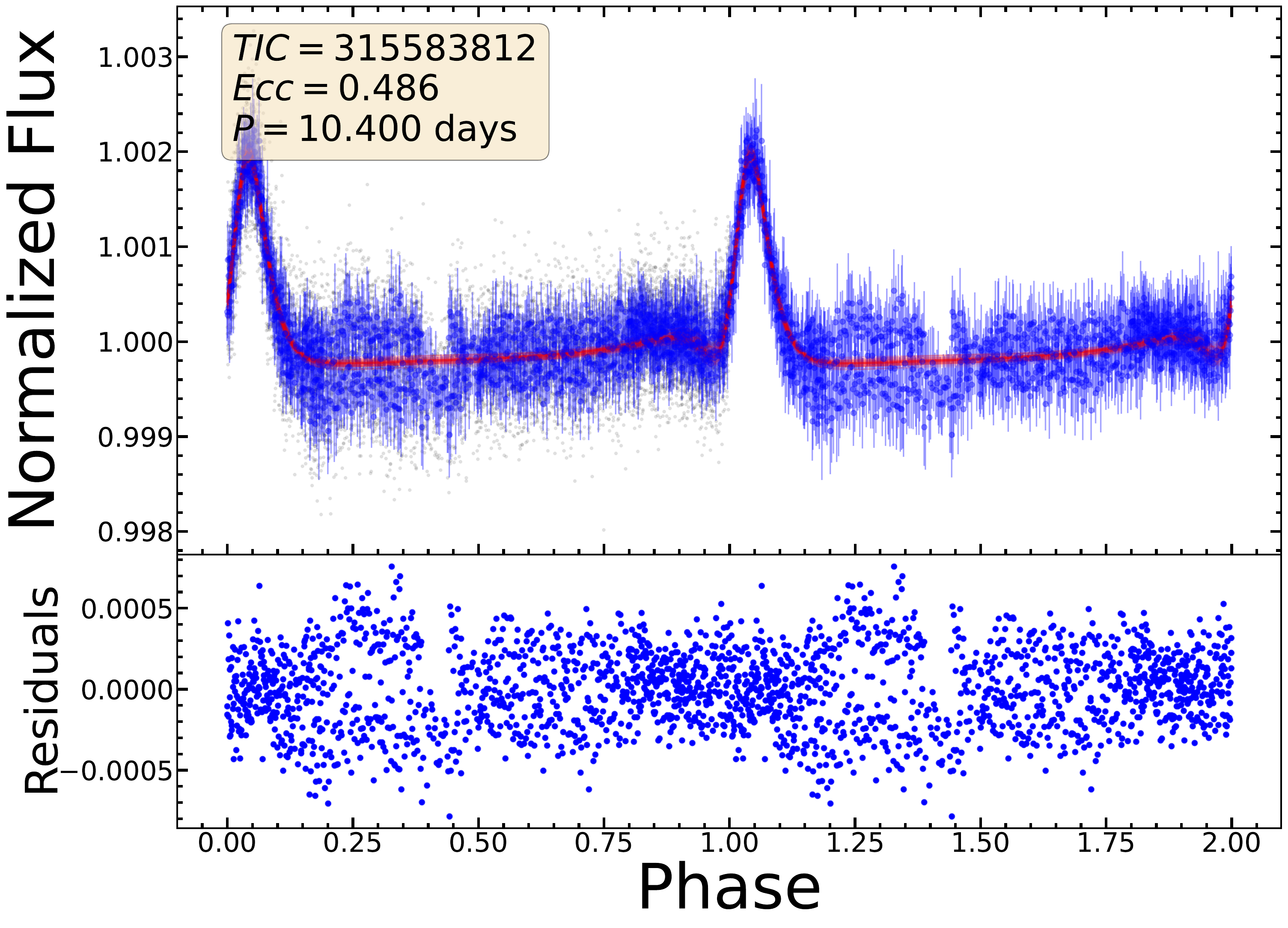} &       
       \includegraphics[width=0.33\textwidth]{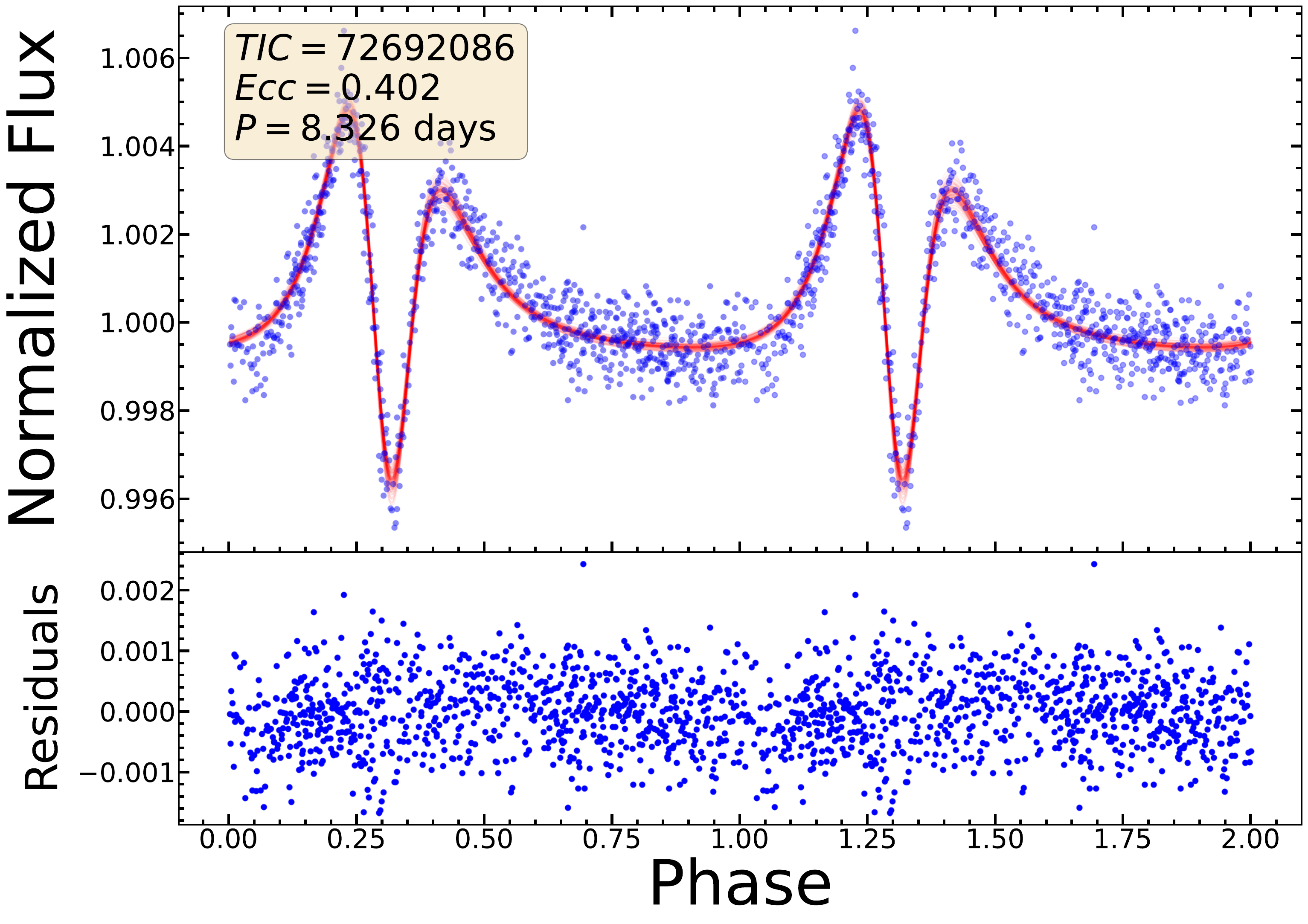} \\

        \includegraphics[width=0.33\textwidth]{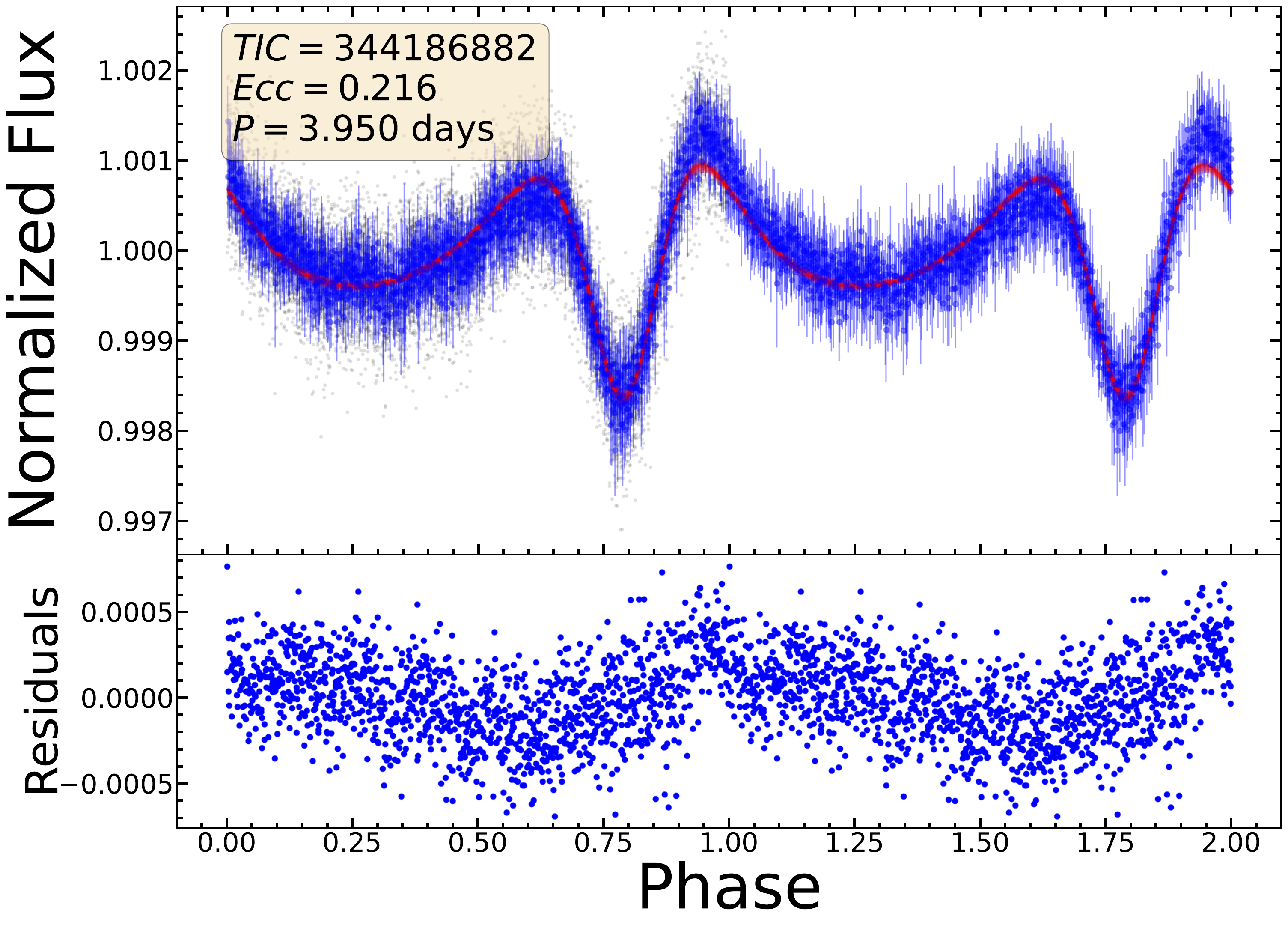} &
       	\includegraphics[width=0.33\textwidth]{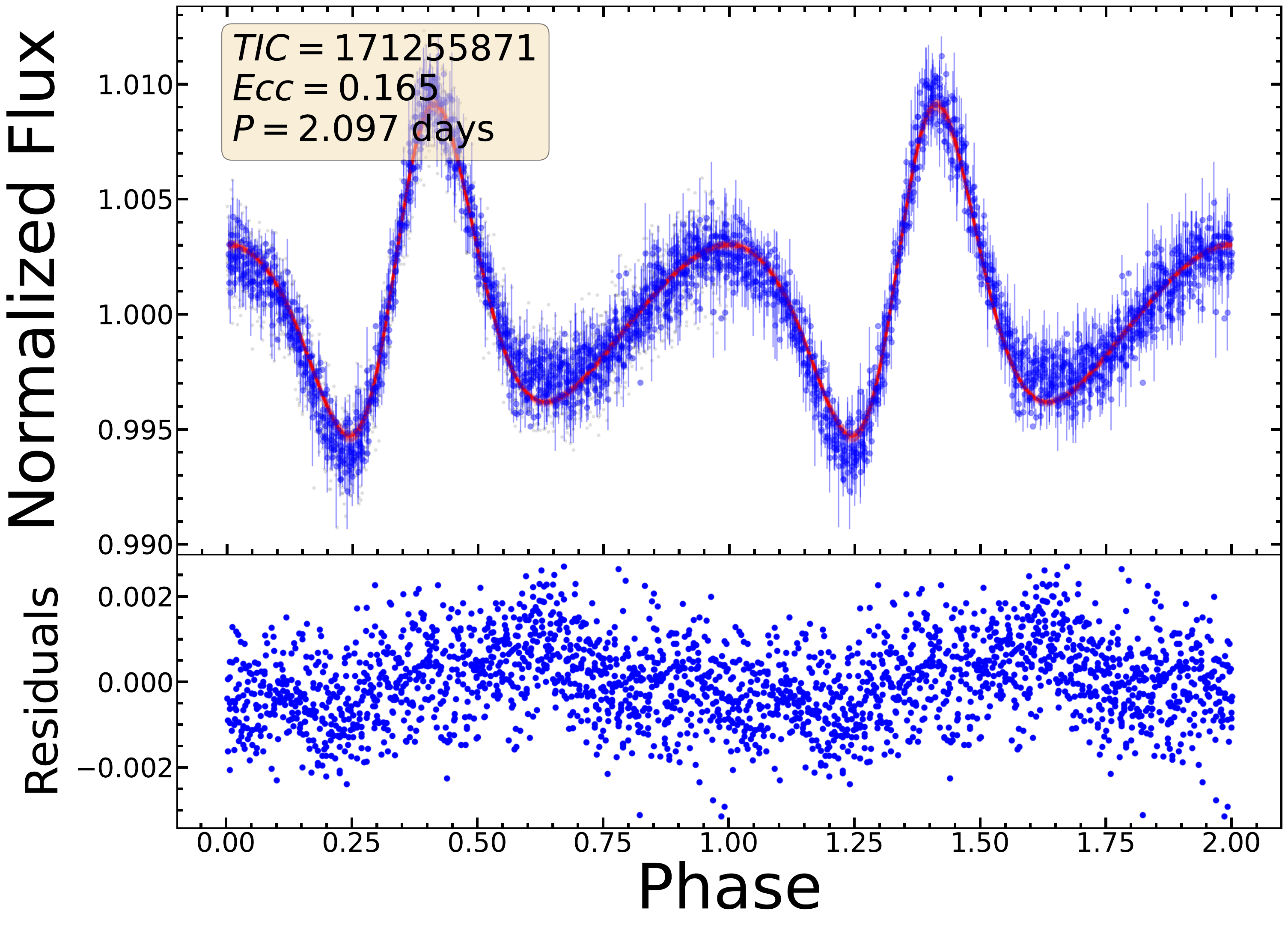} &      
       \includegraphics[width=0.33\textwidth]{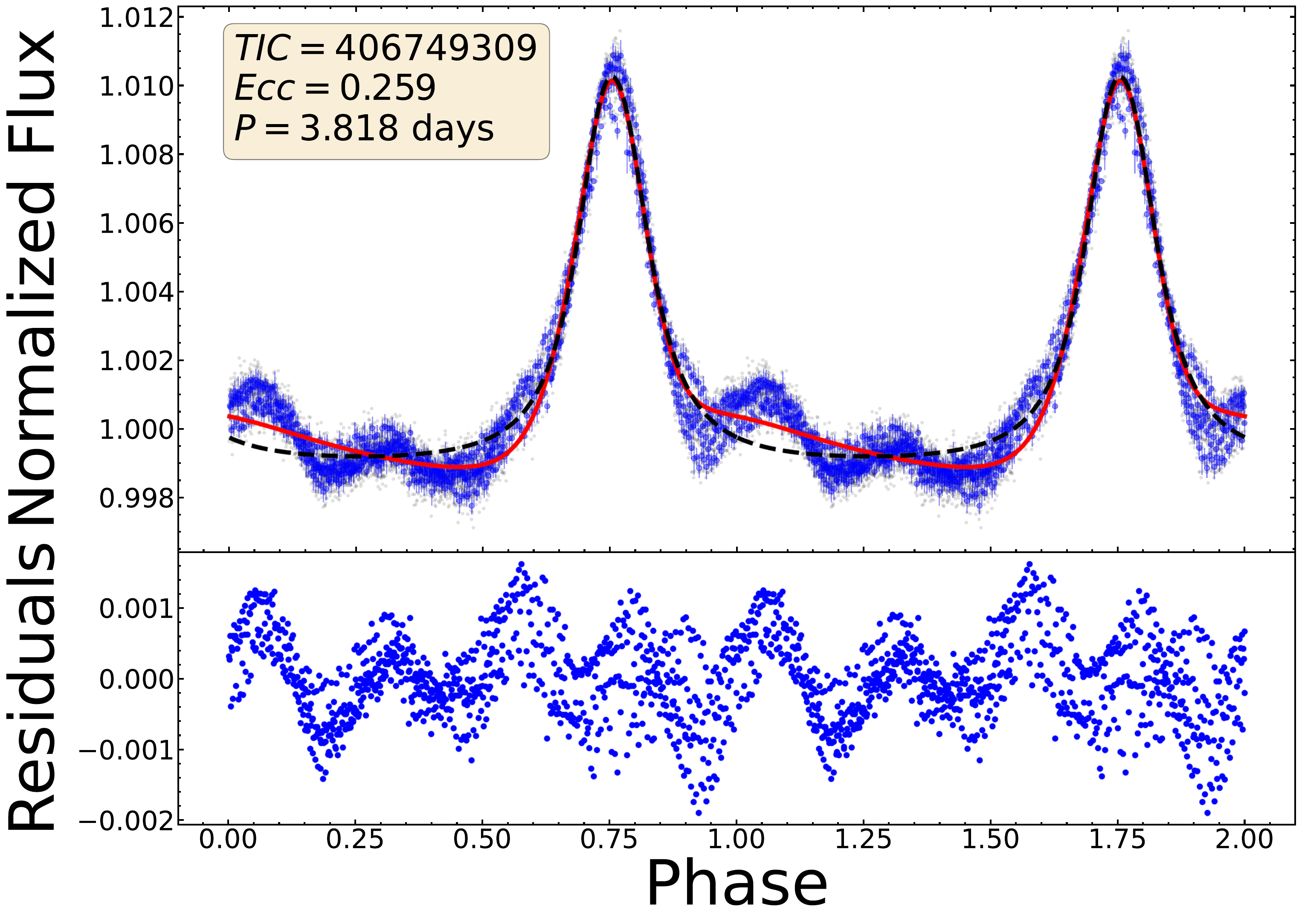} \\

        \includegraphics[width=0.33\textwidth]{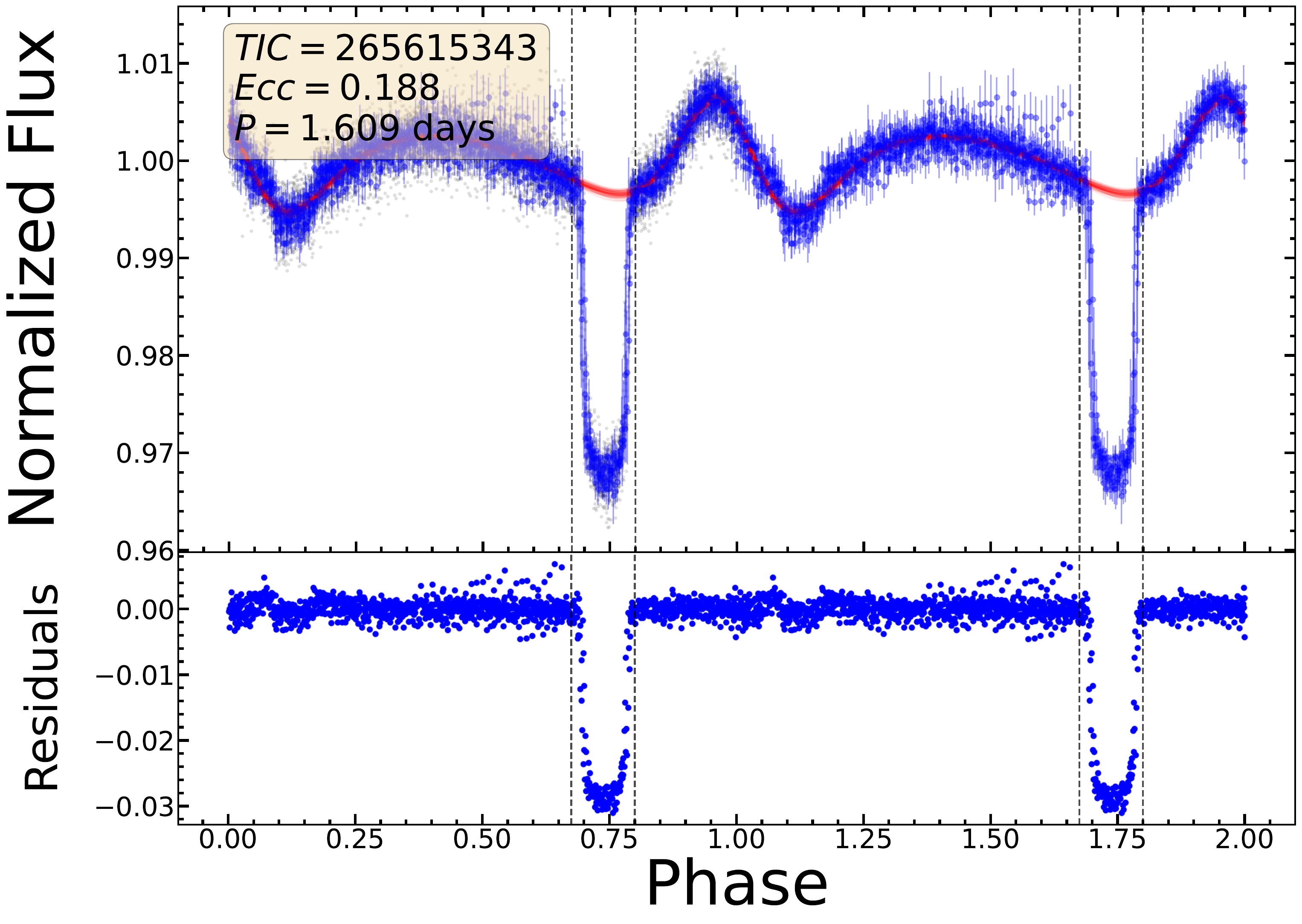} &
	\includegraphics[width=0.33\textwidth]{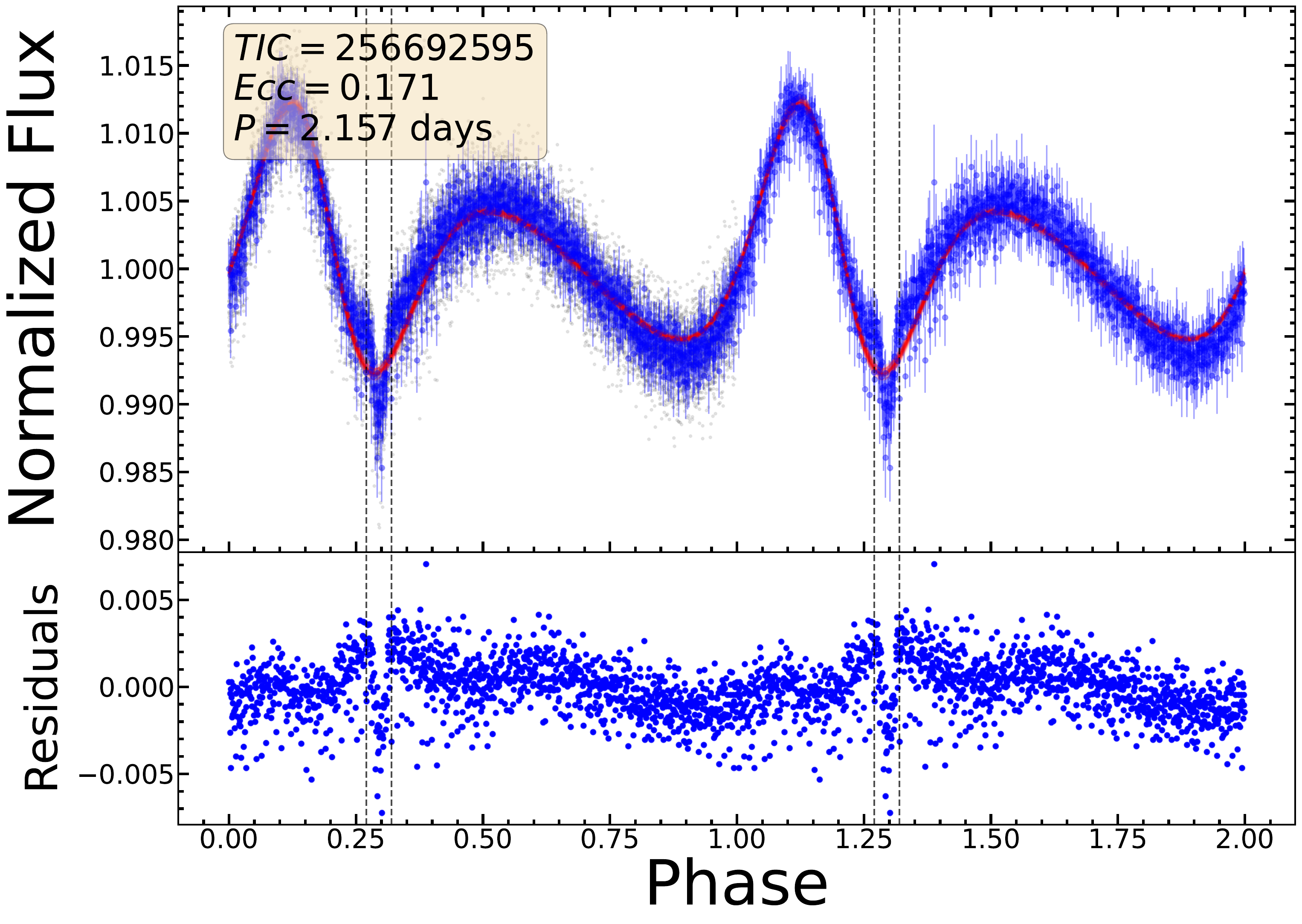} &       
       \includegraphics[width=0.33\textwidth]{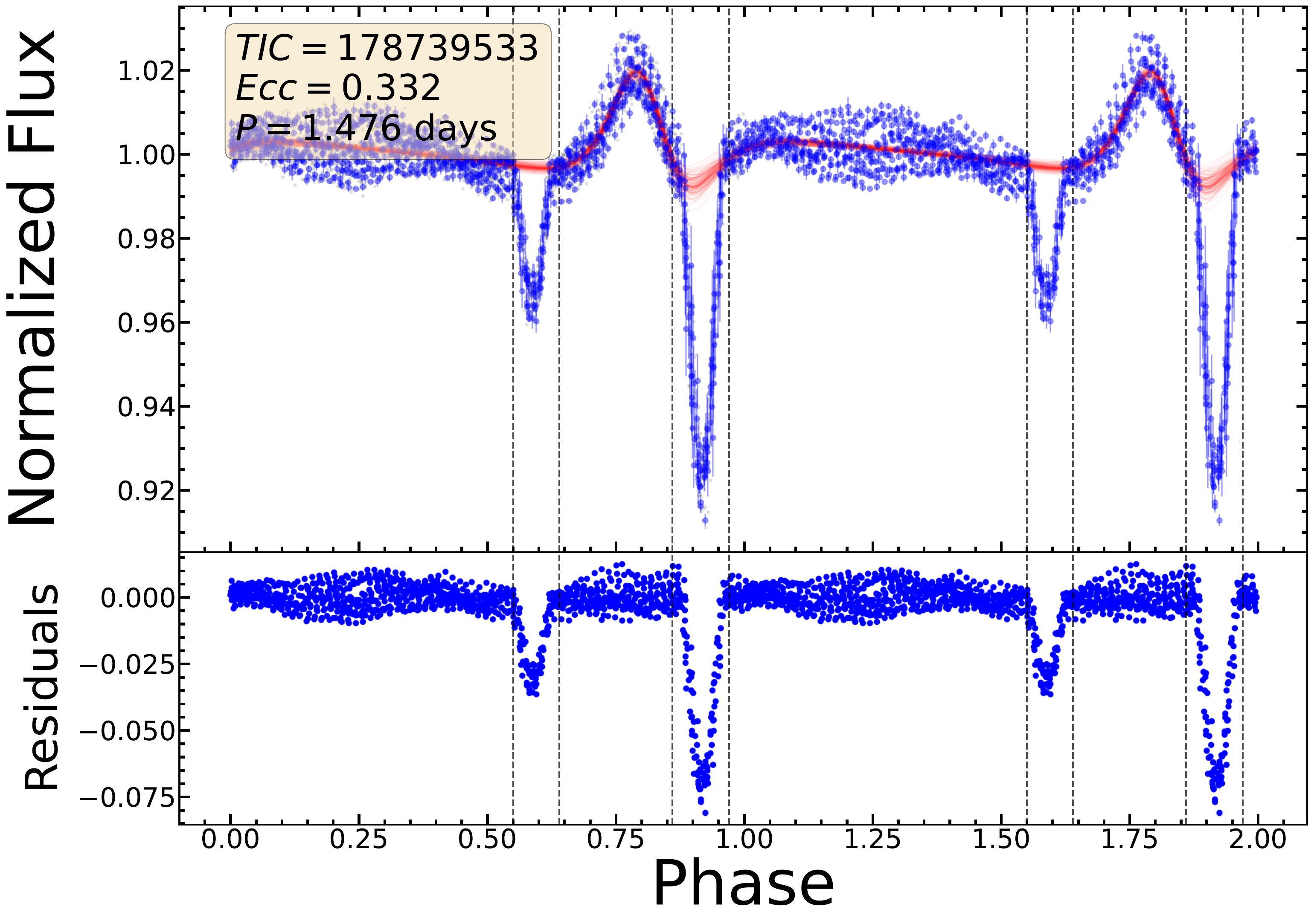} \\
    \end{tabular}
    \caption{Examples of the phase-folded light curves of main sequence APOGEE HBs in half-hour time-bins. The model residuals are shown below each light curve. The \textit{TESS} input catalog (TIC) identifier, the eccentricity, and the period are given in the upper left corner of the light curves. The 406749309 light curve (right column second from the bottom) shows the accepted (solid) solution, consistent with the RV orbit, and the rejected (dashed) solution for this system in Table~\ref{tab:table}. The bottom row shows three HBs that also have eclipses. For these targets, we masked the data in between the vertical dashed lines before deriving final orbital parameters.}
    \label{fig:HB}
\end{figure*}

For each target, we then selected the sector with the lowest $\chi^2_{HB}$ model. To identify HB candidates, we compared the $\chi_{HB}^2$ fit of the HB model against that of a linear fit, $\chi_{line}^2$. Fig.~\ref{fig:Final} shows the distribution of $R=\chi_{HB}^2/\chi_{line}^2$ and orbital eccentricities for the final APOGEE sample. In \citetalias{Me}, we defined the candidate region ($R<0.5$ and $e>0.15$) based on the results of the double-lined spectroscopic (SB2) binary search, which used visual inspection to identify HBs. Here, we have moved the eccentricity limit down to $e>0.10$ to include targets where eclipses have biased the model fit. This left us with 2,940 HB candidates, most of which we reject through visual inspection as either systematic light curve problems or non-HB variables. In total, we identify 50 strong HB candidates. Fig.~\ref{fig:Final} shows the $R$ and $e$ of the selected targets and Fig.~\ref{fig:HB} shows the light curves of fifteen systems. After selecting the HB stars through visual inspection, we model the light curves using Markov Chain Monte Carlo (MCMC) methods as implemented by \citet{FM2013} with 30 walkers for 10,000 iterations removing the first 1000 as burn-in. Table~\ref{tab:table} reports the median posteriors and the $1\sigma$ uncertainty ranges on the eccentricity, inclination, and argument of periastron.

Out of our 50 HBs, 14 have primary and/or secondary eclipses. The bottom row of Fig.~\ref{fig:HB} shows three examples of eclipsing systems. In order to accurately model targets with eclipses, we mask the eclipses and remodel the light curves. In Fig.~\ref{fig:Final} there is one obvious outlier flagged in Table~\ref{tab:table} with an asterisk, TIC 372058935 is below the $e<0.10$ threshold. This is an example of the eclipse masking leading to a change in the eccentricity of the model and subsequently a poorly defined inclination and periastron. We indicate which targets have primary or secondary eclipses in Table~\ref{tab:table}.

\begin{figure*}
    \centering
    \begin{tabular}{cc}
	\includegraphics[width=0.5\textwidth]{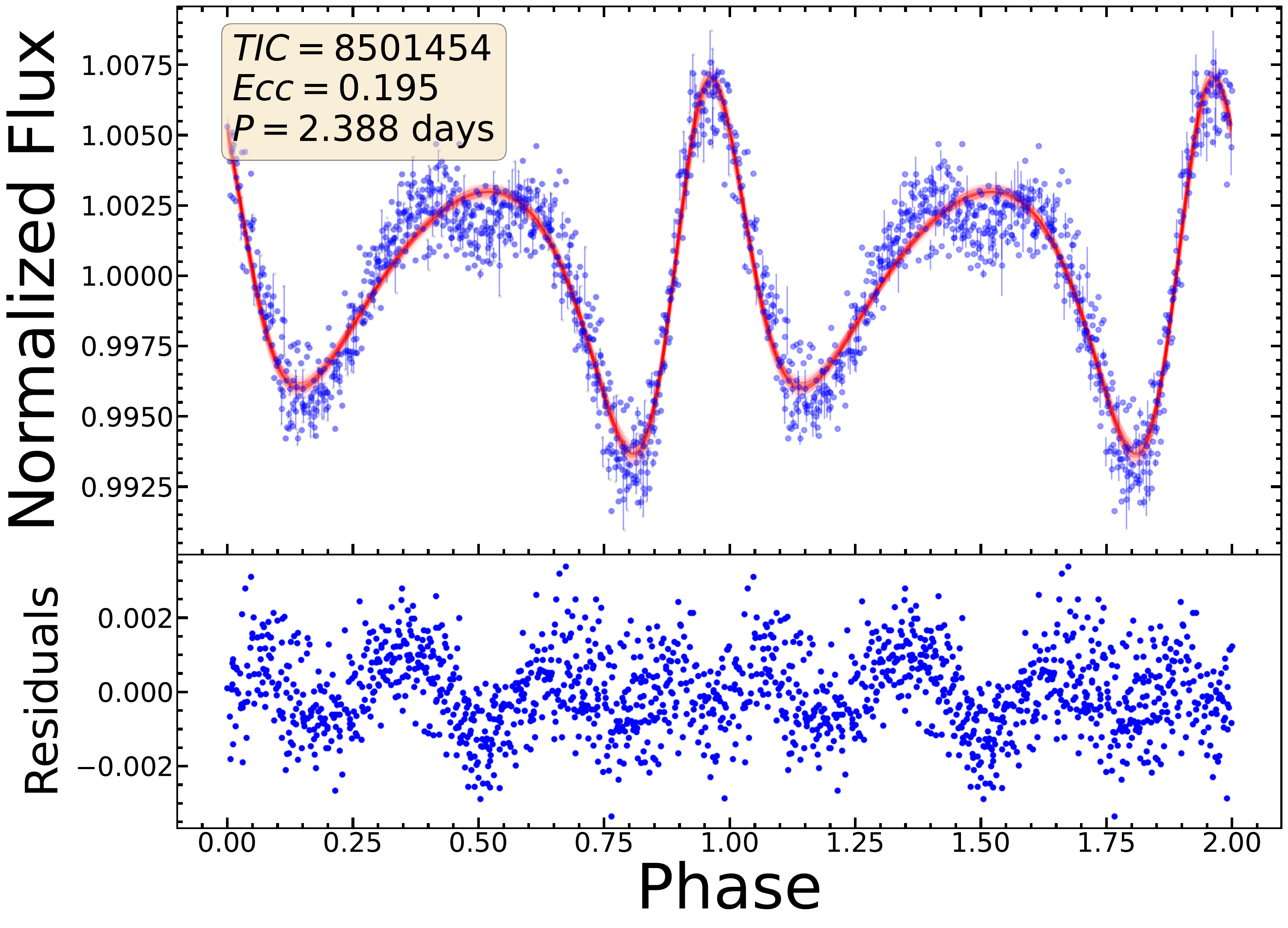} &
    \includegraphics[width=0.5\textwidth]{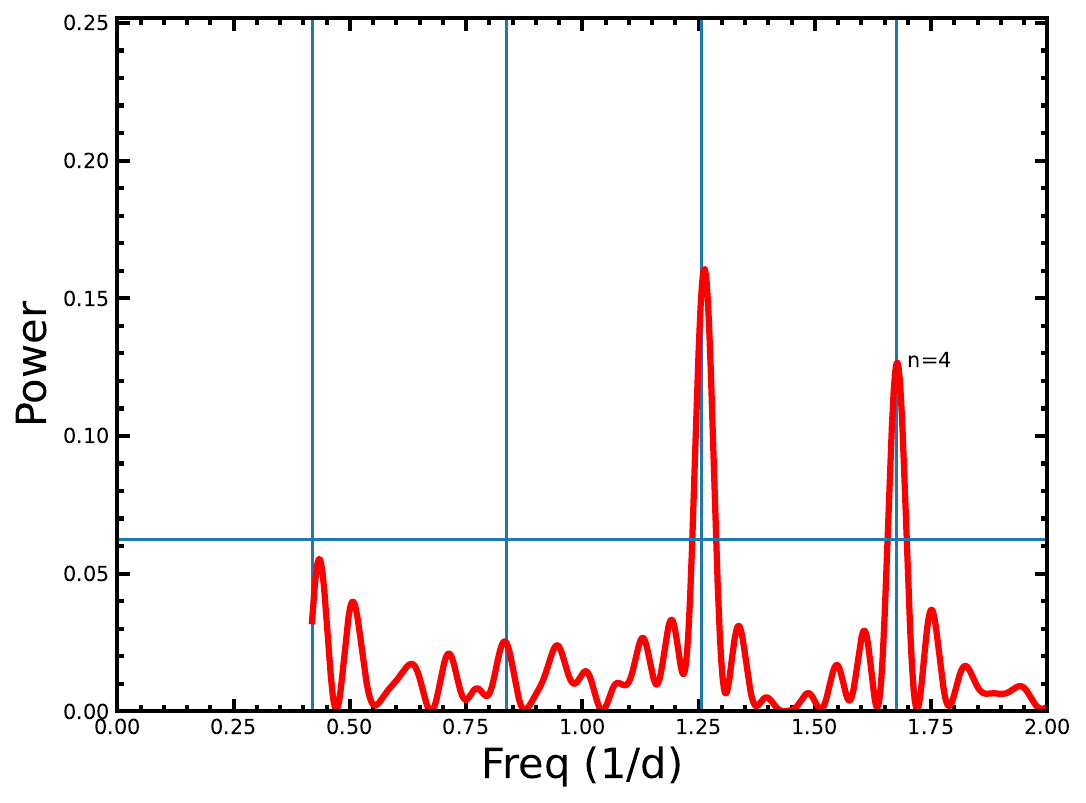} \\
    
    \includegraphics[width=0.5\textwidth]{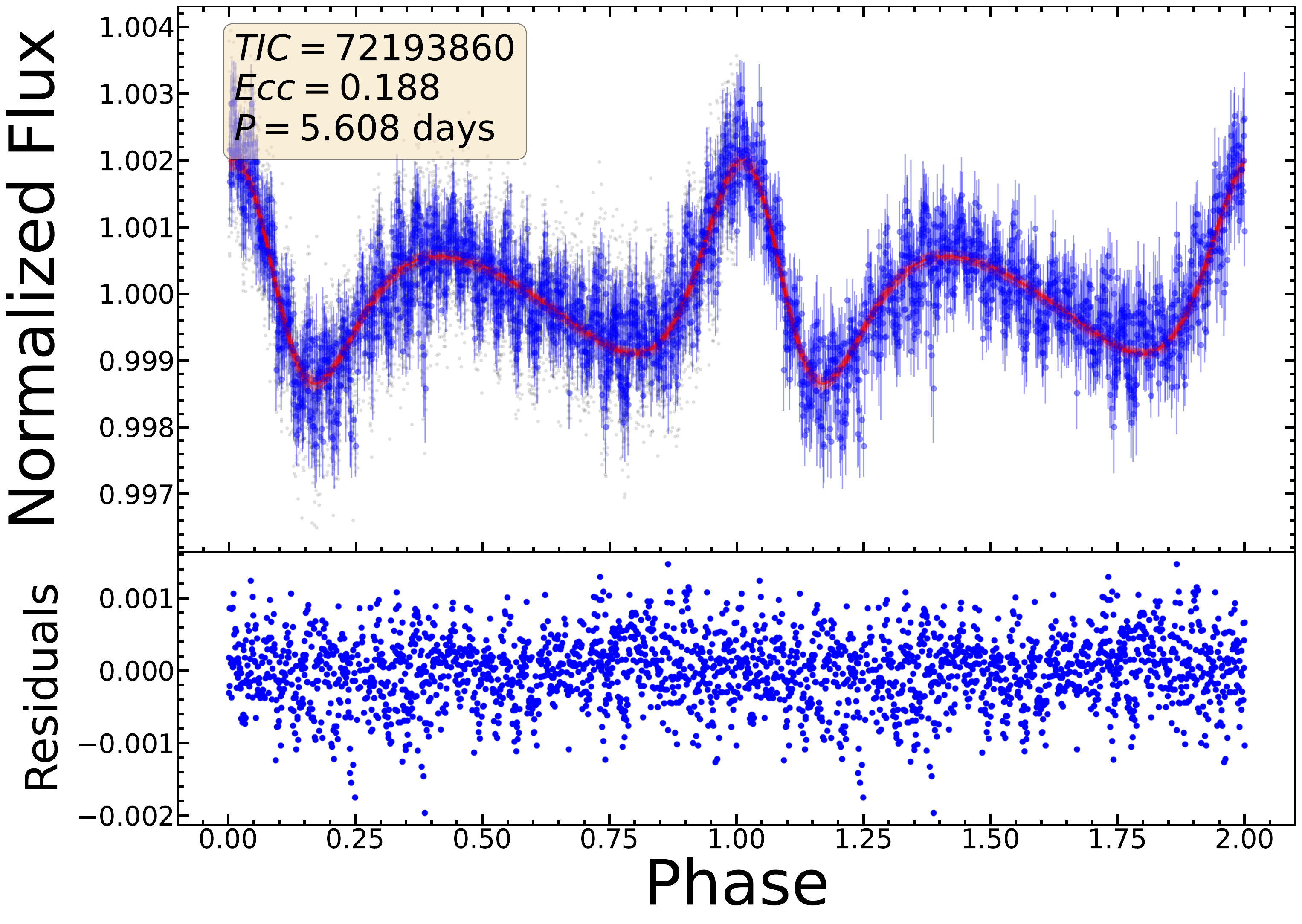} &
    \includegraphics[width=0.5\textwidth]{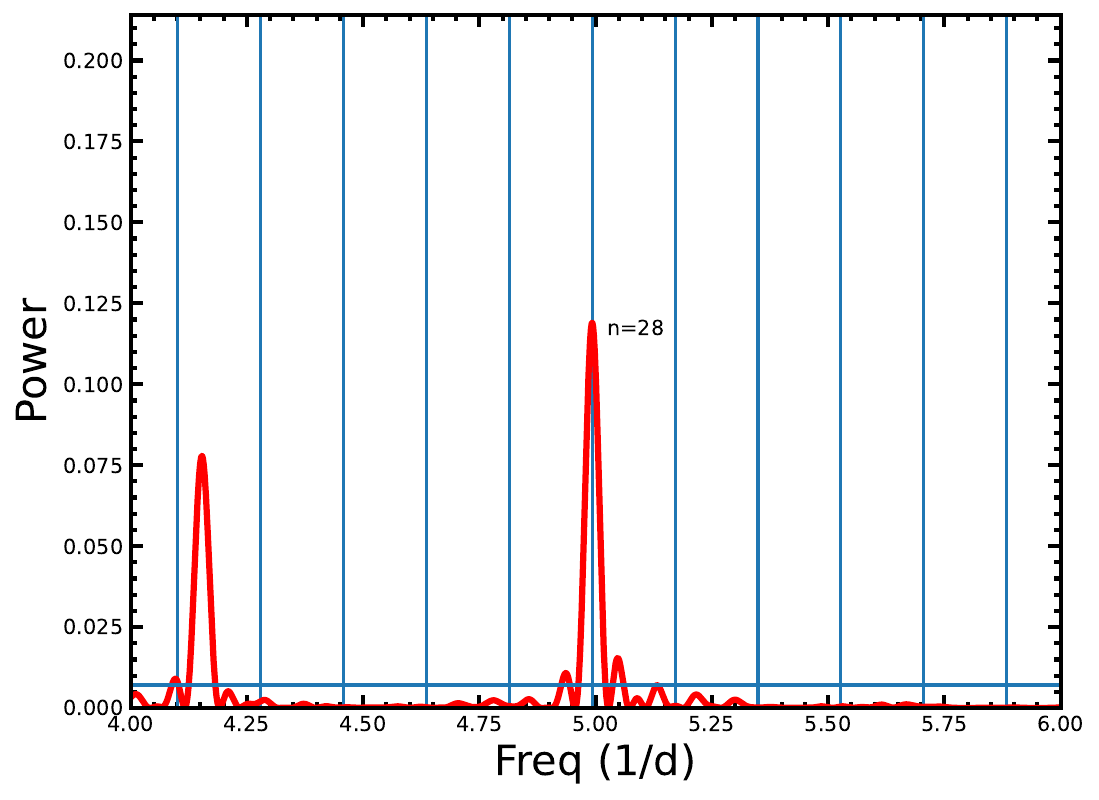} \\
    
    \includegraphics[width=0.5\textwidth]{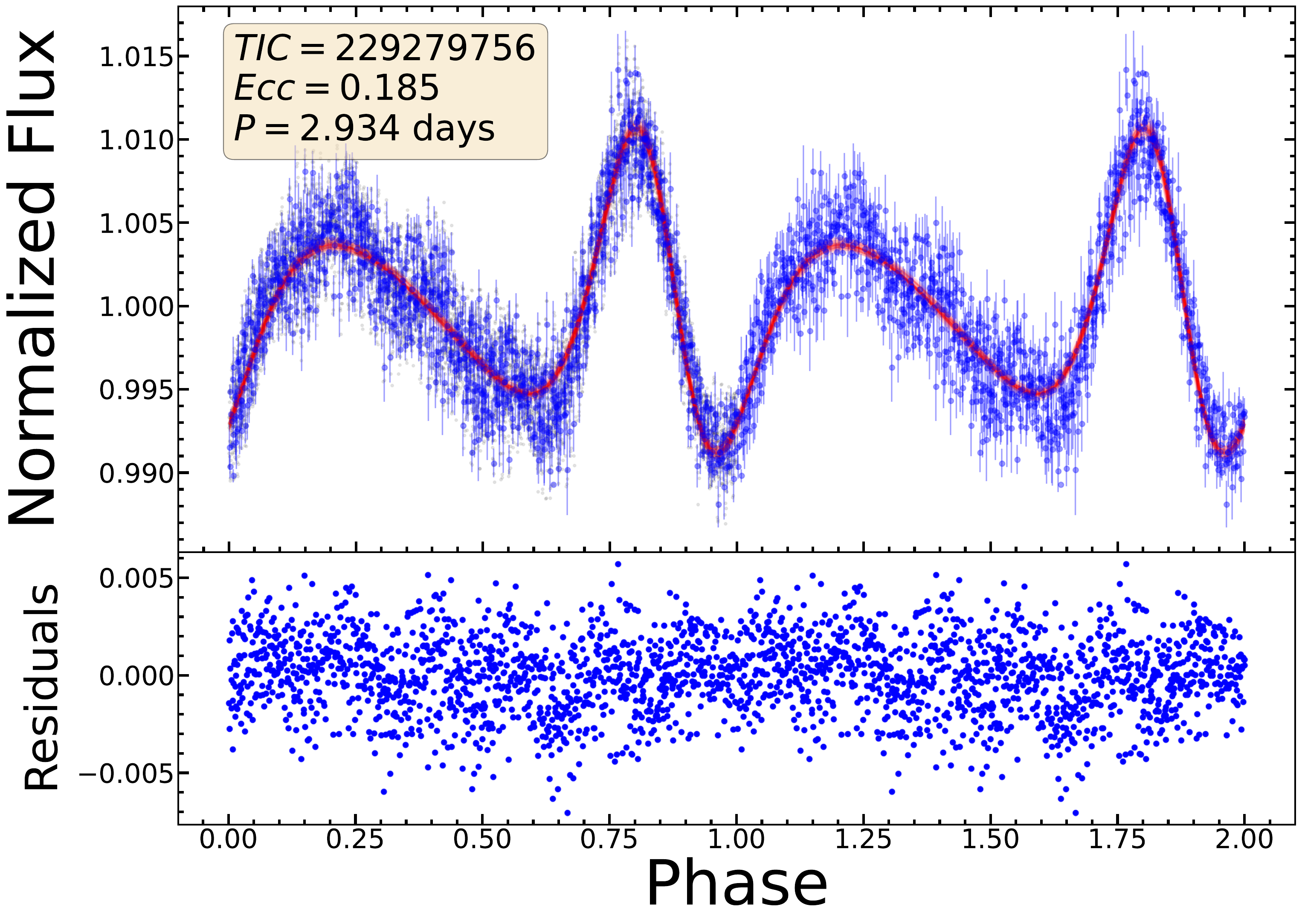} &
    \includegraphics[width=0.5\textwidth]{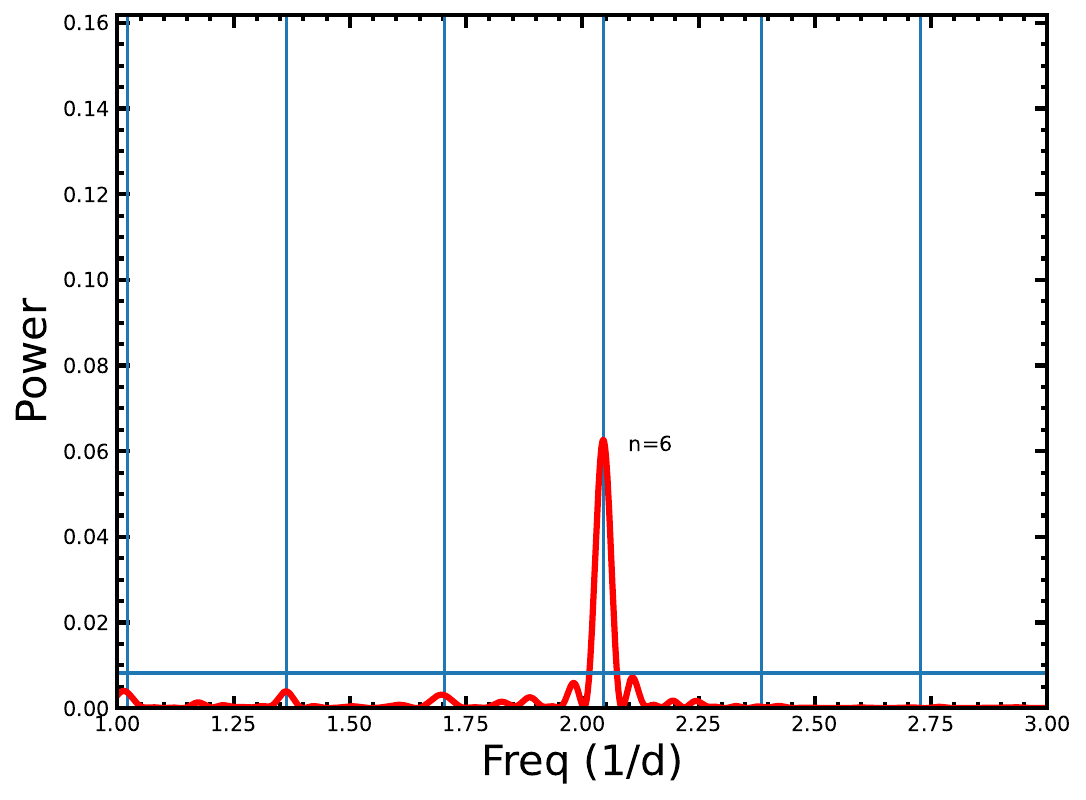} \\

    \end{tabular}
    \caption{Examples of three targets with tidally excited oscillations. The left panels show the phase-folded, binned light curves, the models, and the model residuals. The \textit{TESS} input catalog (TIC) identifier, the eccentricity, and the period are labeled in the upper left of each light curve. The right panels show the periodograms of the residuals with vertical lines at orbital harmonic frequencies ($n/P$), with on harmonic peaks labeled $n=\#$, and a horizontal line at a false alarm probability of $10^{-10}$. The high power off harmonic peaks are examples of non-TEO pulsations.}
    \label{fig:TEO}
\end{figure*}

Additionally, we flag targets with TEOs in Table~\ref{tab:table} to highlight the systems that can be used to study the internal structure of the star. As in \citetalias{Me}, we used periodograms of the residuals from the \citet{Kumar1995} models to search for TEOs. The flagged targets have periodogram powers corresponding to a false alarm probability $< 10^{-10}$ at an integer harmonic $n/P$ of the orbital frequency. Fourteen of the 50 APOGEE HB systems have TEO-like harmonic oscillations, Fig.~\ref{fig:TEO} shows three examples. The median harmonic is $n=5$ with a maximum at $n=28$ for TIC 72193860. The stars can have oscillations at other periods and some examples are seen in Fig.~\ref{fig:TEO}. Finally, Table 1 lists the StarHorse mass estimates \citep{2018Queiroz, 2019Anders, 2022Anders}. While these assume single stars, they should be reasonable estimates for the primary mass unless it is a near-twin binary.

%\subsection{Faint Binary Systems}
%\label{sec:FBS}

%The lower magnitude limit of the \textit{TESS} QLP is $13.5$ magnitudes \citep{Huang2020}. However, we identified 6,705 APOGEE MS binaries observed by \textit{TESS} with $T_{mag}>13.5$ that can not have QLP light curves. In order to create the light curves of these remaining targets we extracted light curves from the full frame images. Following \citet{2023Fausnaugh}, we plotted the targets using $cts\_per\_s - bkg\_model$, where cts\_per\_s is flux counts per seconds and bkg\_model is the model of systematic errors. Additionally, we could not fit an entire phase as a majority of the targets had breaks in the observations and are thrown out. However, when modeling the remaining light curves using the methods of Section~\S\ref{sec:HM} we discovered that the amplitudes of the resulting fluxes are too minimal for accurate fits. In total, we were unable to fit and extract any faint targets for our HB search.

\begin{figure*}
    \begin{tabular}{ccc}
        \includegraphics[width=0.5\textwidth]{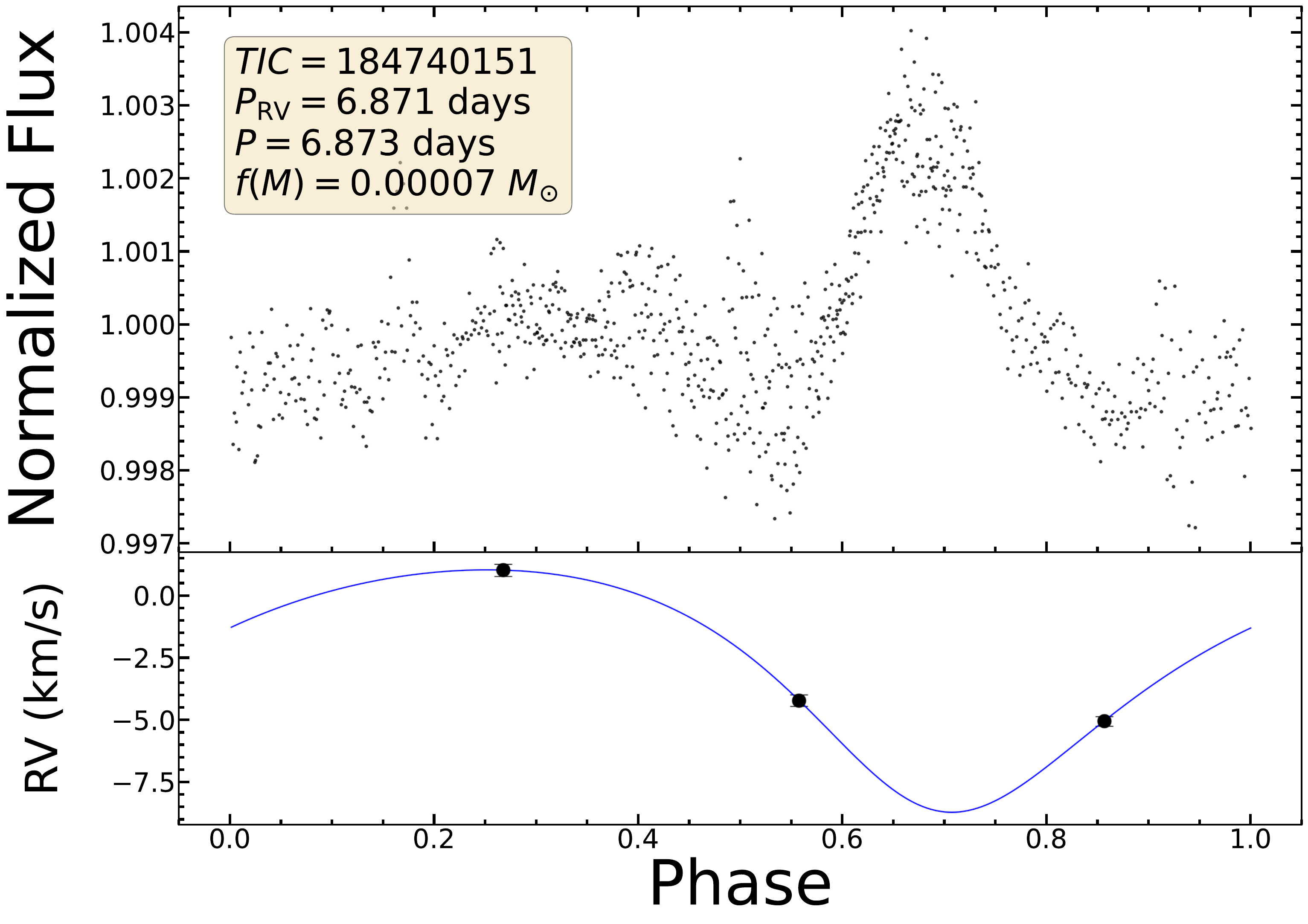} &       
       \includegraphics[width=0.5\textwidth]{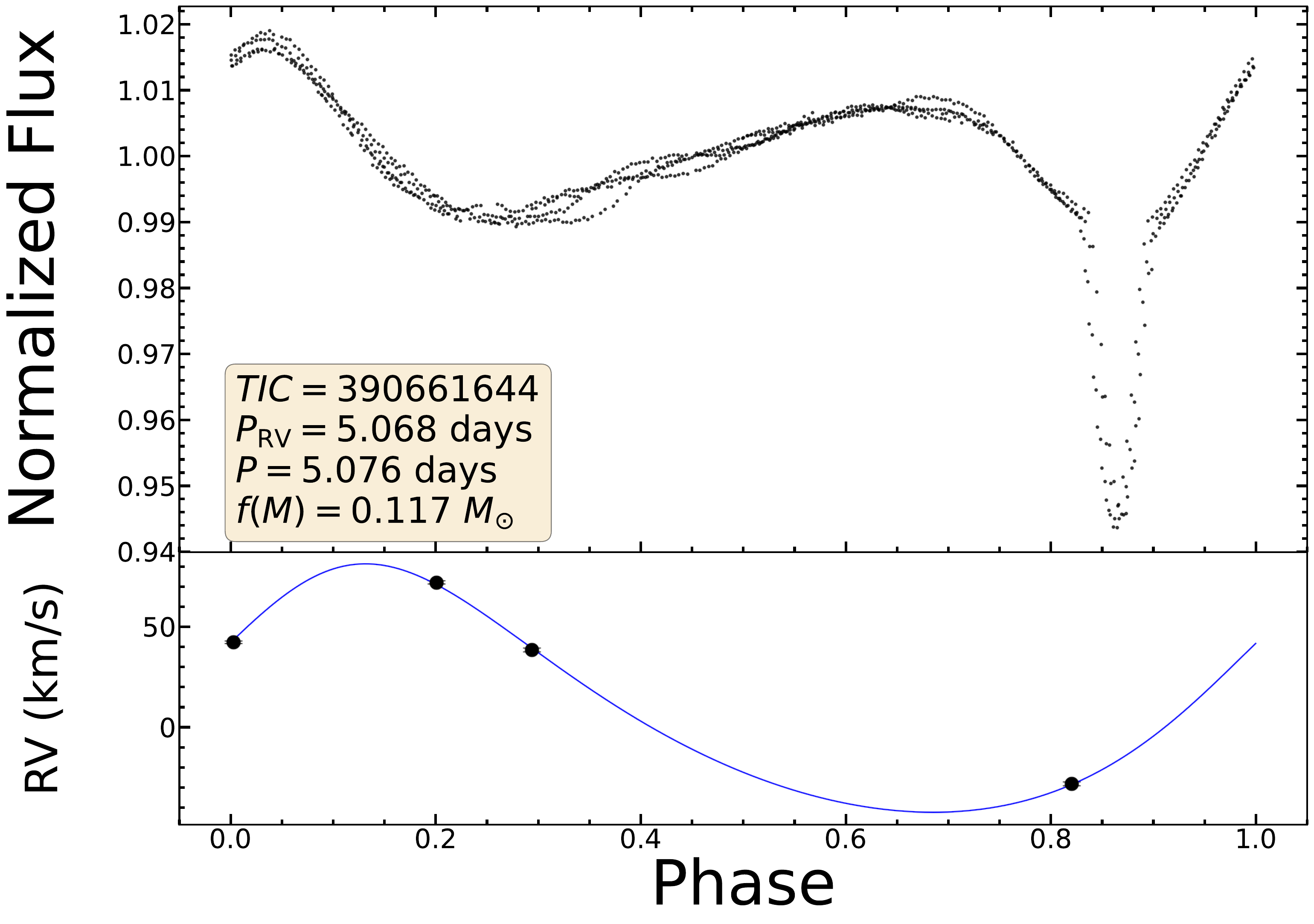} \\
       
        \includegraphics[width=0.5\textwidth]{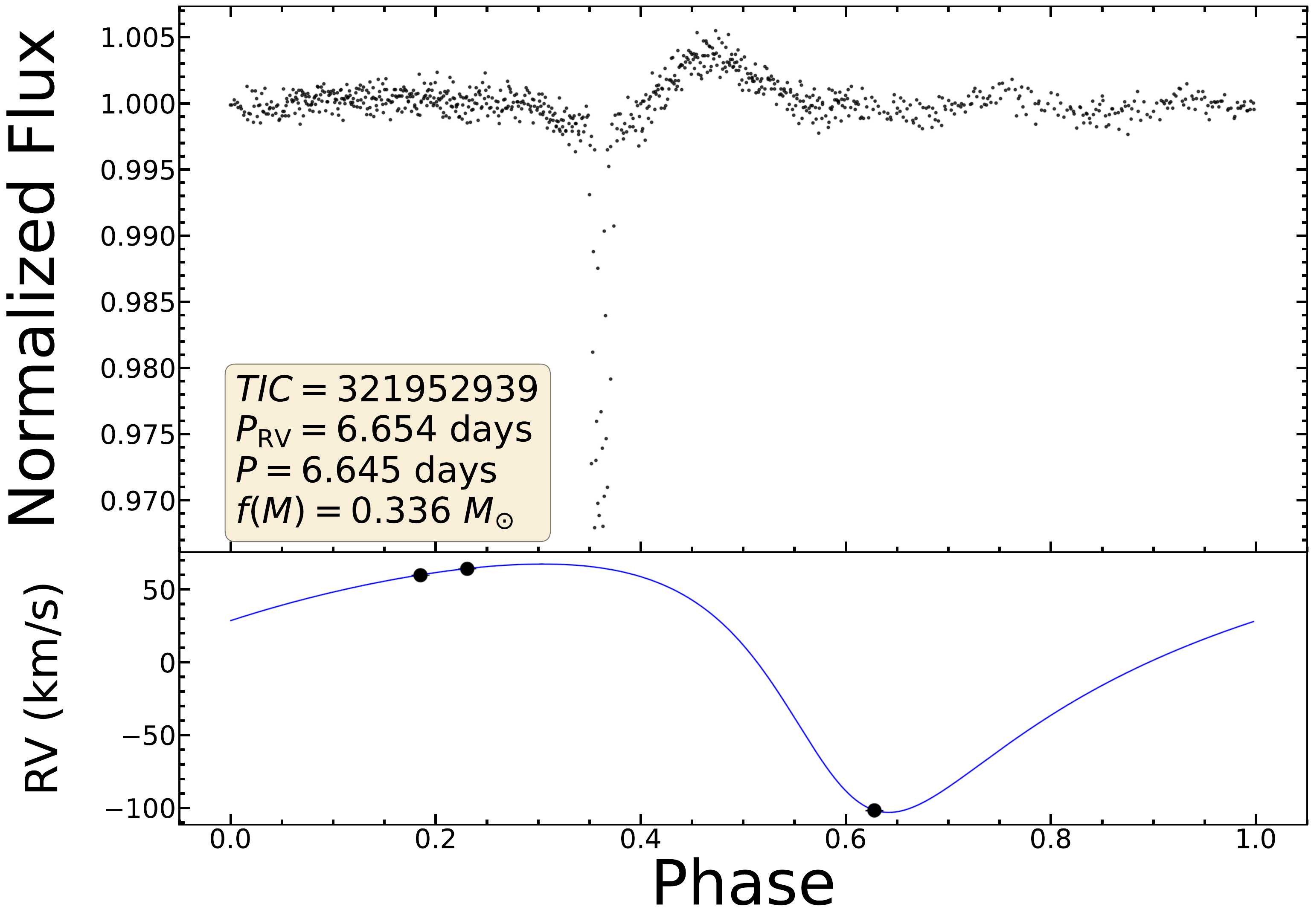} &
       \includegraphics[width=0.5\textwidth]{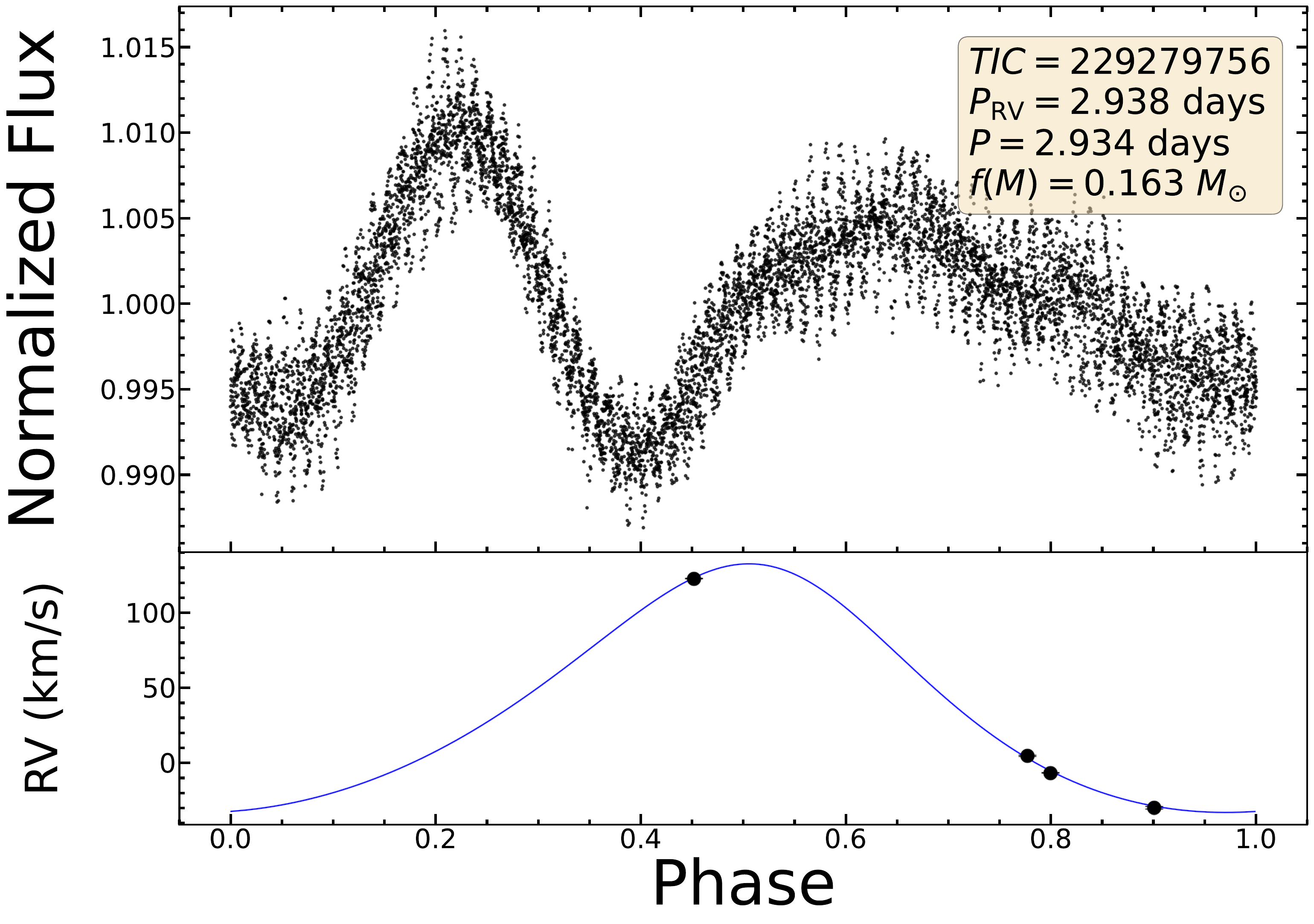} \\

    \end{tabular}
    \caption{Examples of targets with sufficient RV phase coverage to estimate the RV amplitude. The top panels show the phase-folded \textit{TESS} light curves and the bottom panels show the APOGEE RV measurements and the best-fitting RV orbits. }
    \label{fig:RV}
\end{figure*}

\section{Radial Velocities}
\label{sec:RV}

In principle, a key advantage of the APOGEE-based HB search is that APOGEE, unlike \textit{Gaia} DR3, provides the epoch RV measurements. However, many of the RV observations of our 50 HBs are flagged as rejected for containing broad lines or rapid rotation when compared to synthetic template spectra \citep{2015Nidever}. All epochs with these flags are discarded, and only targets with 3 or more remaining epochs of observations are fit with an RV model. Many of the remaining RVs are flagged as "Suspect Broad Lines" even though no RV rejection flags are set. We model the RVs as

\begin{equation}
    RV = {\gamma} + K\bigl(\cos(\omega+\nu)+e\cos\omega\bigr),
	\label{eq:RV}
\end{equation} where the true anomaly, $\nu$ (Eq.~\ref{eq:True}), the argument of periastron, $\omega$, and the eccentricity, $e$, are determined from the light curve. The RV amplitude, $K$, and the center-of-mass velocity, $\gamma$, are free parameters.

Because the APOGEE data are obtained over a much longer time period than a \textit{TESS} sector, the period derived from the \textit{TESS} light curve is likely not accurate enough to properly phase the RV data. To solve this problem, we fit the RV model over a period grid $P-0.02 < P_{\rm{RV}} < P+0.02$ where $P$ is the period from the light curve. We identify the 
$P_{RV}$, $K$, and $\gamma$ that produced the best fit to the RV data. Fig.~\ref{fig:RV} shows a sample of the RV targets. 

In some cases, the RV measurements were not sufficiently spread in orbital phase to constrain the velocity amplitude. To discard the targets with poor phase coverage, we find the distance in phase $D_\pm$ of the RV measurement closest to the phase of the velocity maximum ($+$) and minimum ($-$) and then discard targets with $\max[D_+,D_-] > 0.25$, leaving the 8 targets shown in Table~\ref{tab:table_rv}. From the combined RV and HB model, we determine the mass function

\begin{equation}
    f(M) = {\frac{M_{2}^{3}\sin^3{i}}{(M_{1}+M_{2})^2}} = {\frac{P_{RV}K^{3}}{2{\pi}G}}(1-e^{2})^{3/2},
	\label{eq:BMF}
\end{equation} where $M_{1}$ and $M_2$ are the primary and secondary masses, respectively.

\begin{table*}
    \centering
    \begin{threeparttable}
	   \caption{Properties of the detected heartbeat stars. We report the orbital period, $P$, eccentricity, $e$, inclination, $i$, and argument of periastron, $\omega$, from the light curve fit with 1$\sigma$ uncertainties. We flag which systems have detected eclipses and tidally excited oscillations. For systems with TEOs, $n$ gives the harmonic(s) of the detected TEO(s). We mark which targets have constrained radial velocity orbits reported in Table~\ref{tab:table_rv}. The primary masses, $M_1$, are taken from StarHorse \citep{2018Queiroz, 2019Anders, 2022Anders}. Finally, we include references for targets that were also detected in other \textit{TESS} heartbeat star searches.}
      \renewcommand{\arraystretch}{1.4}
\begin{tabular}{| c  r  c  c  c  c  c  c  c  c  c |}
        \hline
		TIC & \multicolumn{1}{c}{$P$} & $e$ & incl & $\omega$ & Eclipses & TEOs & $n$ & RVs? & $M_{1}$ & Reference\\
          & \multicolumn{1}{c}{(days)} &   & ($\degree$) & ($\degree$) &   &   &   &   & ($M_{\odot}$) &  \\
         \hline

8496206 & 4.098 & $0.356^{+0.016}_{-0.017}$ & $85.04^{+3.48}_{-5.07}$ & $17.17^{+4.56}_{-4.46}$ & \xmark & \xmark & $-$ & \xmark & 1.86 & (4)\\
8501454 & 2.388 & $0.195^{+0.006}_{-0.006}$ & $70.54^{+6.25}_{-4.09}$ & $340.12^{+2.34}_{-2.34}$ & \xmark & \cmark & 4 & \xmark & 2.44 & \xmark\\
13083990 & 7.889 & $0.505^{+0.012}_{-0.012}$ & $32.12^{+0.68}_{-0.73}$ & $229.10^{+2.81}_{-3.06}$ & \xmark & \xmark & $-$ & \xmark & 4.21 & \xmark\\
17373454 & 10.106 & $0.274^{+0.007}_{-0.007}$ & $31.78^{+0.48}_{-0.49}$ & $253.43^{+1.34}_{-1.38}$ & \xmark & \xmark & $-$ & \xmark & 3.96 & \xmark\\
26412885 & 3.490 & $0.154^{+0.006}_{-0.006}$ & $78.37^{+7.38}_{-6.60}$ & $85.66^{+2.38}_{-2.43}$ & \xmark & \xmark & $-$ & \xmark & 2.13 & (3)\\
27762545 & 3.270 & $0.371^{+0.020}_{-0.020}$ & $37.14^{+1.35}_{-1.36}$ & $20.38^{+6.21}_{-6.54}$ & \xmark & \xmark & $-$ & \xmark & 2.62 & \xmark\\
29453664 & 3.847 & $0.251^{+0.008}_{-0.008}$ & $71.14^{+8.30}_{-5.19}$ & $119.28^{+3.30}_{-3.23}$ & \xmark & \xmark & $-$ & \cmark & 4.20 & \xmark\\
50606518 & 6.505 & $0.528^{+0.011}_{-0.011}$ & $48.99^{+1.18}_{-1.08}$ & $71.15^{+3.01}_{-2.98}$ & \xmark & \cmark & 14 & \xmark & 4.43 & \xmark\\
54228807 & 4.045 & $0.319^{+0.007}_{-0.008}$ & $46.86^{+1.13}_{-1.07}$ & $313.23^{+2.41}_{-2.33}$ & \cmark & \xmark & $-$ & \cmark & 4.12 & \xmark\\
72193860 & 5.608 & $0.188^{+0.010}_{-0.010}$ & $45.38^{+1.57}_{-1.44}$ & $30.67^{+3.55}_{-3.61}$ & \xmark & \cmark & 28 & \cmark & 2.62 & \xmark\\
72692086 & 8.326 & $0.402^{+0.007}_{-0.007}$ & $42.44^{+0.55}_{-0.54}$ & $257.26^{+1.71}_{-1.74}$ & \xmark & \xmark & $-$ & \xmark & 3.57 & \xmark\\
74554192 & 10.915 & $0.365^{+0.007}_{-0.007}$ & $47.23^{+0.76}_{-0.73}$ & $254.60^{+1.88}_{-1.88}$ & \xmark & \cmark & 5,6 & \xmark & 2.85 & \xmark\\
80196252 & 2.629 & $0.221^{+0.007}_{-0.007}$ & $31.12^{+0.53}_{-0.54}$ & $298.33^{+1.59}_{-1.53}$ & \xmark & \xmark & $-$ & \xmark & 3.76 & \xmark\\
82355125 & 4.578 & $0.233^{+0.006}_{-0.006}$ & $56.51^{+1.61}_{-1.47}$ & $292.97^{+2.10}_{-2.08}$ & \xmark & \xmark & $-$ & \xmark & 1.75 & \xmark\\
85177749 & 3.900 & $0.282^{+0.012}_{-0.011}$ & $32.45^{+0.68}_{-0.69}$ & $217.98^{+2.67}_{-2.75}$ & \xmark & \xmark & $-$ & \xmark & 1.90 & (1),(3),(4)\\
%105655404 & 5.832 & $0.290^{+0.014}_{-0.015}$ & $34.36^{+0.86}_{-0.87}$ & $286.38^{+3.06}_{-3.03}$ & \xmark & \cmark & 1,5 & \xmark & 2.70 & \xmark\\
136040527 & 3.205 & $0.352^{+0.018}_{-0.024}$ & $60.08^{+12.44}_{-11.61}$ & $273.71^{+3.40}_{-3.12}$ & \cmark & \xmark & $-$ & \xmark & 2.27 & \xmark\\
140810845 & 3.546 & $0.281^{+0.002}_{-0.002}$ & $39.52^{+0.20}_{-0.20}$ & $211.01^{+0.73}_{-0.73}$ & \xmark & \cmark & 5,8 & \xmark & 2.43 & \xmark\\
%150257543 & 13.701 & $0.106^{+0.009}_{-0.009}$ & $26.04^{+0.79}_{-0.84}$ & $0.19^{+0.33}_{-0.15}$ & \xmark & \cmark & 2,6,8 & \xmark & 1.93 & \xmark\\
171255871 & 2.097 & $0.165^{+0.004}_{-0.004}$ & $43.39^{+0.66}_{-0.62}$ & $149.42^{+1.67}_{-1.63}$ & \xmark & \xmark & $-$ & \xmark & 2.06 & (4)\\
172594580 & 10.061 & $0.441^{+0.028}_{-0.022}$ & $32.83^{+1.36}_{-1.63}$ & $184.34^{+8.13}_{-7.20}$ & \xmark & \xmark & $-$ & \cmark & 2.04 & \xmark\\
178739533 & 1.476 & $0.332^{+0.021}_{-0.019}$ & $42.74^{+2.44}_{-2.37}$ & $31.84^{+5.47}_{-5.65}$ & \cmark & \xmark & $-$ & \xmark & 3.03 & (2)\\
184740151 & 6.873 & $0.252^{+0.016}_{-0.016}$ & $34.79^{+1.09}_{-1.08}$ & $324.07^{+4.48}_{-4.21}$ & \xmark & \xmark & $-$ & \cmark & 2.92 & \xmark\\
229279756 & 2.934 & $0.185^{+0.007}_{-0.006}$ & $50.49^{+1.54}_{-1.42}$ & $213.66^{+2.76}_{-2.76}$ & \xmark & \cmark & 6 & \cmark & 1.94 & \xmark\\
241344367 & 3.909 & $0.202^{+0.011}_{-0.012}$ & $57.05^{+4.15}_{-3.26}$ & $6.68^{+4.95}_{-3.76}$ & \xmark & \xmark & $-$ & \xmark & 2.15 & (3),(4)\\
247560986 & 5.403 & $0.381^{+0.010}_{-0.010}$ & $39.77^{+0.78}_{-0.78}$ & $12.90^{+3.15}_{-3.31}$ & \xmark & \xmark & $-$ & \xmark & 4.22 & \xmark\\
256692595 & 2.157 & $0.171^{+0.005}_{-0.005}$ & $42.75^{+0.73}_{-0.70}$ & $219.24^{+1.72}_{-1.73}$ & \cmark & \xmark & $-$ & \xmark & 2.74 & \xmark\\
265615343 & 1.609 & $0.188^{+0.009}_{-0.010}$ & $60.27^{+4.30}_{-3.34}$ & $200.53^{+4.10}_{-3.97}$ & \cmark & \xmark & $-$ & \xmark & 1.93 & \xmark\\
266071175 & 3.396 & $0.419^{+0.008}_{-0.008}$ & $35.53^{+0.48}_{-0.49}$ & $250.58^{+1.71}_{-1.78}$ & \xmark & \xmark & $-$ & \xmark & 2.40 & \xmark\\
269295421 & 2.546 & $0.234^{+0.022}_{-0.019}$ & $27.30^{+1.09}_{-1.15}$ & $321.56^{+4.18}_{-4.05}$ & \cmark & \xmark & $-$ & \xmark & 3.87 & \xmark\\
279060522 & 4.537 & $0.500^{+0.006}_{-0.006}$ & $32.83^{+0.27}_{-0.27}$ & $223.26^{+1.20}_{-1.20}$ & \xmark & \cmark & 3,8 & \xmark & 5.38 & \xmark\\
281316097 & 2.270 & $0.182^{+0.011}_{-0.011}$ & $26.27^{+0.58}_{-0.62}$ & $31.38^{+2.14}_{-2.19}$ & \xmark & \xmark & $-$ & \xmark & 4.95 & \xmark\\
308066884 & 6.917 & $0.588^{+0.006}_{-0.006}$ & $43.20^{+0.54}_{-0.51}$ & $302.09^{+2.14}_{-2.12}$ & \xmark & \xmark & $-$ & \xmark & 1.99 & \xmark\\
315583812 & 10.400 & $0.486^{+0.016}_{-0.015}$ & $33.60^{+1.07}_{-1.18}$ & $147.01^{+5.45}_{-4.92}$ & \xmark & \xmark & $-$ & \xmark & 7.13 & \xmark\\
317766125 & 6.364 & $0.197^{+0.005}_{-0.005}$ & $45.45^{+0.78}_{-0.75}$ & $24.84^{+1.91}_{-1.94}$ & \xmark & \cmark & 2,5 & \xmark & 1.92 & \xmark\\
321952939 & 6.645 & $0.381^{+0.031}_{-0.029}$ & $40.54^{+2.78}_{-2.77}$ & $2.09^{+3.28}_{-1.56}$ & \cmark & \xmark & $-$ & \cmark & 1.83 & (2)\\
344186882 & 3.950 & $0.216^{+0.005}_{-0.005}$ & $59.63^{+2.02}_{-1.74}$ & $278.31^{+2.05}_{-2.05}$ & \xmark & \cmark & 7 & \xmark & 2.04 & (3)\\
348508349 & 4.152 & $0.306^{+0.004}_{-0.004}$ & $45.96^{+0.51}_{-0.49}$ & $209.34^{+1.28}_{-1.29}$ & \cmark & \cmark & 2,4,8,10,11 & \xmark & 6.02 & (2)\\
354959477 & 1.773 & $0.159^{+0.004}_{-0.004}$ & $86.43^{+2.50}_{-3.65}$ & $113.29^{+1.47}_{-1.49}$ & \xmark & \xmark & $-$ & \xmark & 2.59 & \xmark\\
357371372 & 8.799 & $0.383^{+0.015}_{-0.015}$ & $35.08^{+0.94}_{-0.98}$ & $288.84^{+3.41}_{-3.13}$ & \xmark & \xmark & $-$ & \xmark & 1.55 & \xmark\\
359019914 & 1.930 & $0.148^{+0.011}_{-0.011}$ & $55.46^{+5.19}_{-3.95}$ & $207.30^{+6.01}_{-5.99}$ & \cmark & \xmark & $-$ & \xmark & 1.64 & \xmark\\
372058935$^*$ & 3.662 & $0.085^{+0.013}_{-0.041}$ & $84.83^{+3.93}_{-65.31}$ & $181.09^{+95.23}_{-95.76}$ & \cmark & \cmark & 3 & \xmark & 2.30 & \xmark\\
390661644 & 5.076 & $0.185^{+0.004}_{-0.004}$ & $49.14^{+0.94}_{-0.93}$ & $313.66^{+1.49}_{-1.50}$ & \cmark & \xmark & $-$ & \cmark & 4.94 & \xmark\\
391085159 & 4.827 & $0.276^{+0.009}_{-0.009}$ & $41.45^{+0.83}_{-0.78}$ & $30.01^{+2.59}_{-2.66}$ & \cmark & \xmark & $-$ & \xmark & 3.73 & (2)\\
%391187591 & 8.226 & $0.431^{+0.019}_{-0.018}$ & $45.23^{+2.99}_{-2.17}$ & $40.74^{+9.52}_{-11.51}$ & \xmark & \cmark & 4 & \xmark & 6.67 & \xmark\\
399516112 & 7.297 & $0.151^{+0.012}_{-0.012}$ & $81.27^{+6.11}_{-8.04}$ & $296.87^{+4.33}_{-4.29}$ & \xmark & \xmark & $-$ & \xmark & 1.76 & \xmark\\
406749309 & 3.818 & $0.259^{+0.008}_{-0.008}$ & $23.50^{+0.35}_{-0.36}$ & $22.90^{+117.80}_{-3.71}$ & \xmark & \cmark & 4 & \xmark & 11.0 & \xmark\\
406749309$^{**}$ & 3.818 & $0.490^{+0.016}_{-0.034}$ & $15.00^{+2.33}_{-4.95}$ & $175.76^{+93.65}_{-91.61}$ & \xmark & \cmark & 4 & \xmark & 11.0 & \xmark\\
418183908 & 8.858 & $0.565^{+0.016}_{-0.016}$ & $27.51^{+1.32}_{-1.63}$ & $320.19^{+6.23}_{-5.13}$ & \xmark & \xmark & $-$ & \xmark & $-$ & (2)\\
435905890 & 6.438 & $0.317^{+0.011}_{-0.011}$ & $44.59^{+1.47}_{-1.43}$ & $339.83^{+5.95}_{-5.63}$ & \xmark & \xmark & $-$ & \xmark & 2.11 & (4)\\
444446485 & 3.405 & $0.310^{+0.017}_{-0.017}$ & $37.69^{+1.32}_{-1.31}$ & $15.73^{+5.60}_{-5.65}$ & \xmark & \xmark & $-$ & \xmark & 1.78 & \xmark\\
445628275 & 6.824 & $0.297^{+0.028}_{-0.031}$ & $37.67^{+3.39}_{-3.44}$ & $61.19^{+5.36}_{-5.69}$ & \cmark & \xmark & $-$ & \xmark & 5.42 & \xmark\\
452366964 & 6.377 & $0.452^{+0.024}_{-0.024}$ & $45.20^{+3.87}_{-3.65}$ & $244.52^{+6.90}_{-7.73}$ & \cmark & \xmark & $-$ & \xmark & 4.03 & (2)\\
468721314 & 4.283 & $0.577^{+0.003}_{-0.003}$ & $30.91^{+0.22}_{-0.23}$ & $127.56^{+0.86}_{-0.84}$ & \xmark & \cmark & 2,3,4,5,6,7 & \xmark & 2.64 & (2),(4)\\

\hline
\end{tabular}
	   \label{tab:table}
    \begin{tablenotes}
      \small
      \item (1) \citet{Li2024a}; (2) \citet{Sol2025}; (3) \citet{Me}; (4) \citet{2025Zhou}
      \item * After masking the eclipses, we find that the eccentricity is below our $e > 0.1$ threshold (Fig.~\ref{fig:Final}).
      \item ** We find two possible solutions for TIC 406749309, and we reject this one on the basis of archival RVs (see \S\ref{sec:RV})
    \end{tablenotes}
  \end{threeparttable}
\end{table*}

\begin{figure}
	\includegraphics[width=\columnwidth]{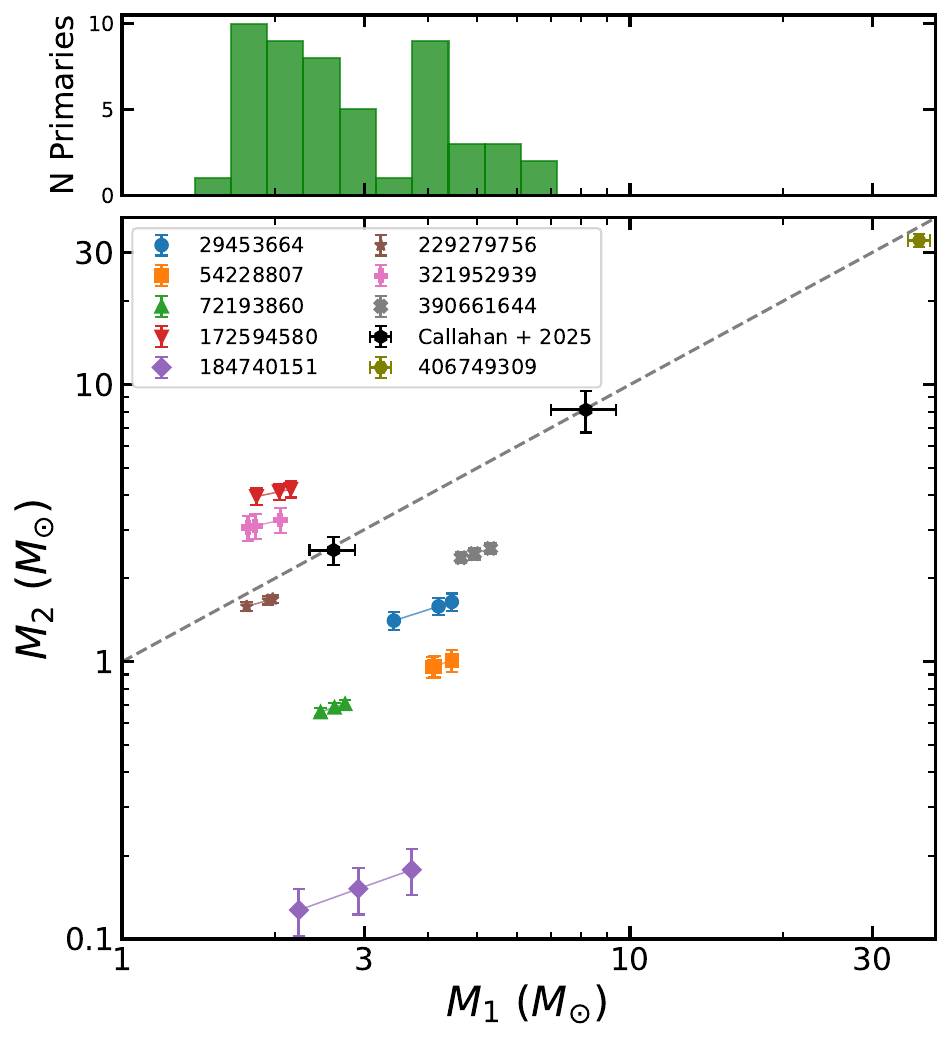}
    \caption{Estimates of the primary and secondary masses. For each system with a mass function measurement reported in Table~\ref{tab:table_rv}, the three points show the secondary mass estimates for the median and 16th/84th percentile estimates of the primary mass \citep{2019Anders, 2022Anders}. The black hexagons are the results from \citet{Me} for two double-lined spectroscopic binary systems The olive hexagon is the result of the mass components from \citet{2012Trepl164816} with the inclination of the accepted solution in Table~\ref{tab:table}. The grey line denotes an equal-mass binary. The histogram is the distribution of StarHorse mass estimates for all the HB primaries \citep{2018Queiroz, 2019Anders, 2022Anders}.}
    \label{fig:Mass}
\end{figure}

As shown in Table~\ref{tab:table}, the typical primary stars are fairly massive, as expected from their luminosities (see \S\ref{sec:D}). For the stars with sufficient phase coverage and reliable mass function estimates, we can also estimate the masses of the secondaries. We use primary masses from StarHorse \citep{2018Queiroz, 2019Anders, 2022Anders}, which reports the median of the mass posterior and the 16th and 84th percentiles. The StarHorse mass may overestimate the true primary mass because it is derived from the spectral energy distribution, which includes a contribution from the secondary. However, because of the steepness of the MS mass-luminosity relation, $L \propto M^\alpha$ with $\alpha \sim 3$ to $4$, it is only a 20-25\% overestimate even for a twin binary. We solve Eq.~\ref{eq:BMF} for the secondary mass, sampling over the uncertainties in the mass function and the inclination. Table~\ref{tab:table_rv} reports these secondary mass estimates, and the results are shown in Fig.~\ref{fig:Mass}. Six out of the 8 systems have $M_2 < M_1$. We highlight the two systems with $M_2 > M_1$ in Table \ref{tab:table_rv} with an asterisk. This is not due to overestimated primary masses from StarHorse, since we would have to increase $M_1$ to find solutions with $M_2 < M_1$, not decrease it. Instead, these are most likely systems where the secondary is substantially affecting the APOGEE velocity estimates by adding luminous spectral features.

One target has archival RVs and a previously reported spectroscopic orbit. TIC 406749309 (HD 164816), is a very bright (7.04 mag) non-eclipsing source with the largest StarHorse mass estimate \citep[$M_1 = 11.0 M_{\odot}$, ][]{2018Queiroz, 2019Anders, 2022Anders}. Our MCMC run found the two solutions shown in Table~\ref{tab:table}. The system is a previously known SB2 with a period, 3.819 days, and eccentricity, 0.232, \citep{2012Trepl164816, 2014Sana164816}, which matches best with our lower $e\simeq0.259$ eccentricity solution, so we adopt it as the true solution shown in Fig.~\ref{fig:HB}. We also show the light curve of the other solution to illustrate the level of degeneracy. The inclination is the lowest of our sample $i\simeq23.50^{\circ}$. \citet{2012Trepl164816} reports masses of $M_1\sin^3{i} = 2.355 \pm 0.069 M_{\odot}$ and $M_2\sin^3{i} = 2.103 \pm 0.062 M_{\odot}$ and rotational velocities of $v_1\sin{i} = 85.4 \pm 0.028 (km/s)$ and $v_2\sin{i} = 79.9 \pm 0.032 (km/s)$. In combination with our inclination estimate, this system consists of two rapidly rotating ($v_1 = (214.2 \pm 3.0)$~km/s and $v_2 = (200.4 \pm 2.8)$~km/s) massive stars ($M_1 = (37.1 \pm 1.9)~M_{\odot}$ and $M_2 = (33.2 \pm 1.7)~M_{\odot}$). These masses are also included in Fig.~\ref{fig:Mass}.

\section{Discussion}
\label{sec:D}

\begin{figure}
	\includegraphics[width=\columnwidth]{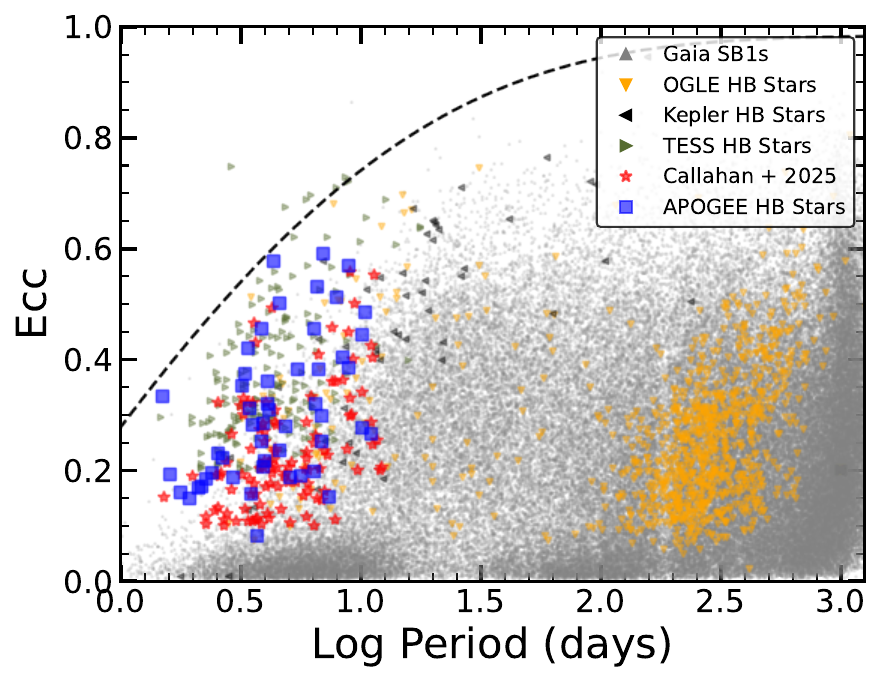}
    \caption{The distribution of orbital periods and eccentricities for the detected APOGEE HBs (squares). The stars show the HBs identified from the Gaia spectroscopic binaries catalogs in \citetalias{Me}. The grey background are the Gaia SB1s after applying the quality cut from \citet{Bashi2022}. The coloured triangles are from the \citet{Wrona2022a, Wrona2022b}, \citet{2023Min-Yu}, and \citet{Sol2025} catalogs of HBs. The dashed curve is the envelope outside of which orbits will rapidly circularize \citep{Mazeh2008}.}
    \label{fig:evp}
\end{figure}

We used the \textit{TESS} light curves of APOGEE MS binaries to identify 50 HBs among the 31,548 MS binaries in the APOGEE data releases 17 and 19 with $T<13.5$. Of these, 14 show eclipses, 14 have TEOs, and 8 have reliable RV fits to determine mass functions. Of our 50 APOGEE HB targets, 14 were previously known across \citet{Li2024a}, \citet{Sol2025}, \citet{Me}, and \citet{2025Zhou}, leaving 36 new HBs. The orbital periods range from 1.48 to 10.92 days, with a median of 4.13 days. The eccentricities range from $e=0.085$ (TIC 372058935) to 0.588 (TIC 308066884), with a median of 0.281. All of our targets fall below the period-eccentricity envelope expected from tidal circularization \citep{Mazeh2008}, as shown in Fig.~\ref{fig:evp}. Although outside the scope of this work, future studies of the stellar dynamics of such unique binary systems would be of great interest.

\begin{figure}
	\includegraphics[width=\columnwidth]{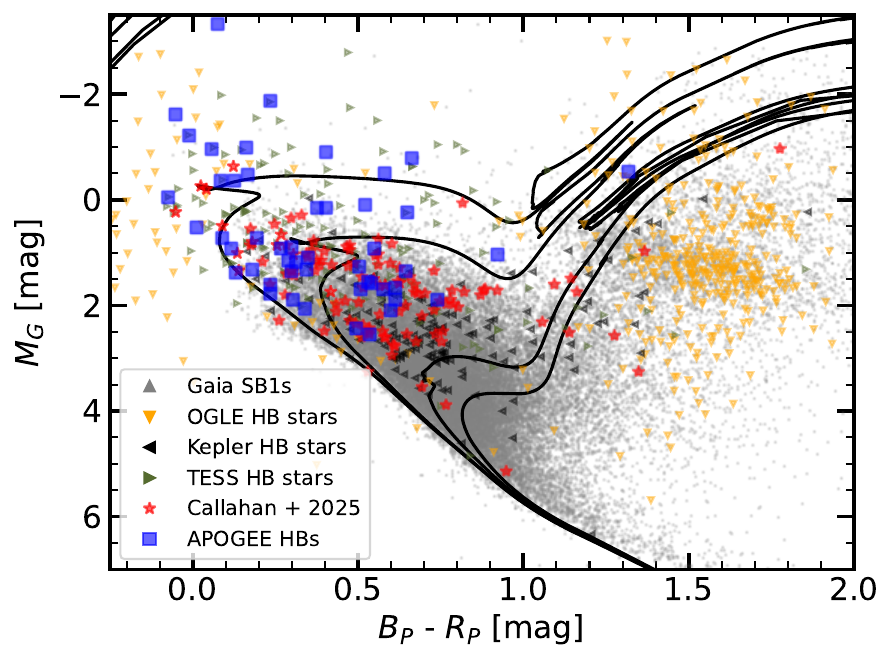}
    \caption{Distribution of HBs on the \Gaia{} colour-magnitude diagram (CMD). The blue squares are the APOGEE HBs detected here. The red stars are the HBs detected in \citetalias{Me} from the \Gaia{} SB1 and SB2 catalogs. The orange, black, and green triangles show HBs from other catalogs \citep{Wrona2022a, Wrona2022b, Shporer2016, 2023Min-Yu, 2017Dimitrov, Kirk2016, Sol2025}. For the OGLE Magellanic Cloud HBs, we used the \citet{2021Skowron} reddening map and the distance and extinction assumptions from \citet{2025MacLeod}. The 0.5, 1, 5, 10 Gyr isochrones are solar metallicity MIST isochrones \citep{2016Dotter, 2016Choi}.}
    \label{fig:CMD}
\end{figure}

\begin{figure}
    \begin{tabular}{c}      
       \includegraphics[width=\columnwidth]{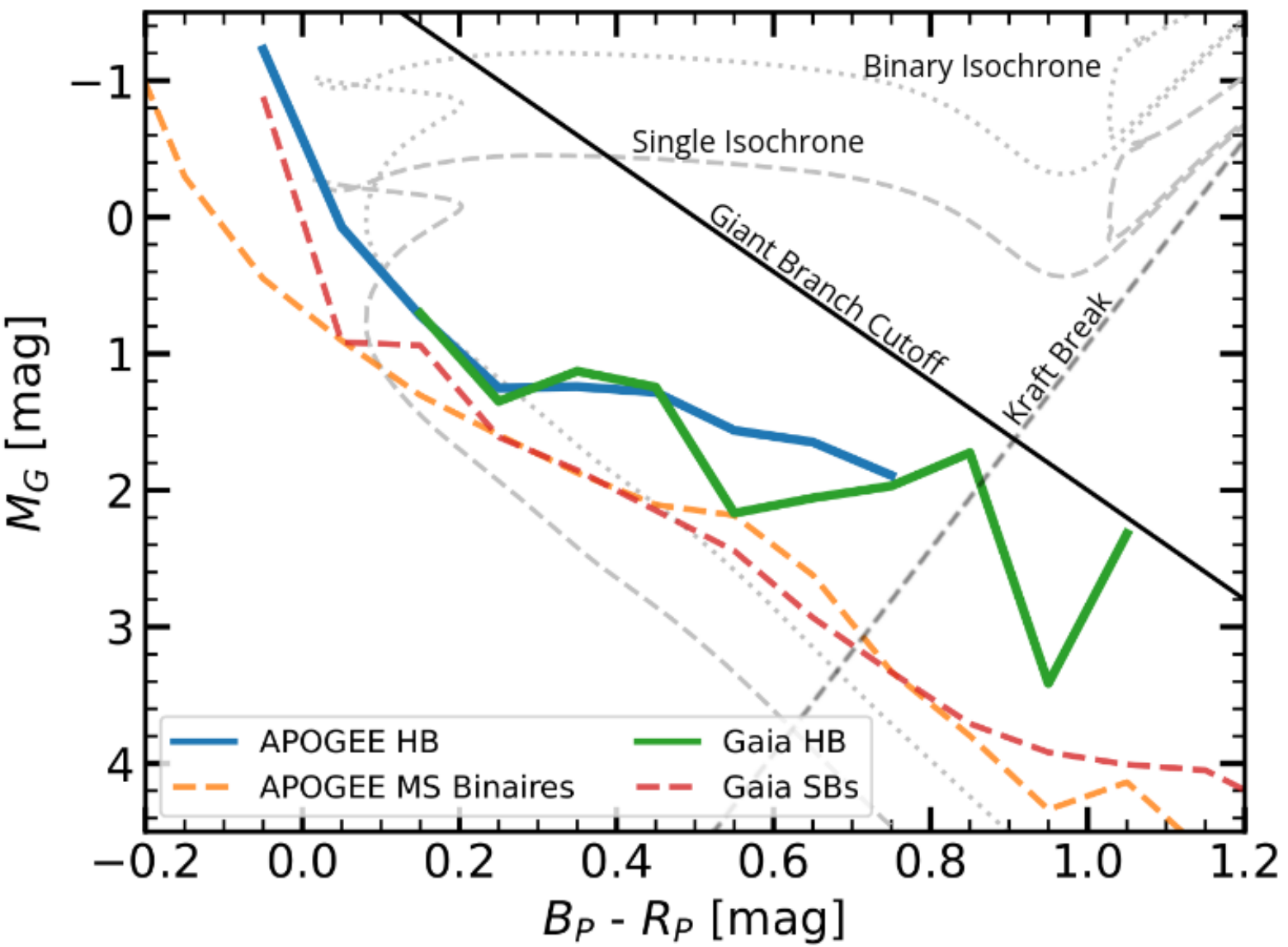} \\
    \end{tabular}
    \caption{Mean magnitudes in 0.1~mag wide colour bins for stars below the giant branch cutoff (black line). The APOGEE HBs binned in colour are shown in blue, and the parent sample of main sequence binaries from APOGEE DR17/19 is shown in orange. Similarly, the binned results from the HBs detected in \citetalias{Me} are shown in green, and the parent sample of \Gaia{} single-lined spectroscopic binaries on the main sequence is shown in red. To guide the eye, a 0.5 Gyr MIST isochrone \citep{2016Dotter, 2016Choi} is shown along with an equal-mass binary isochrone that is brighter by a factor of two, and the Kraft break, which crosses the isochrone at $T_{\rm{eff}} = 6550$~K \citep{1967Kraft, 1962Schatzman, 2024Beyer} to split the upper and lower main sequence.}
    \label{fig:Line}
\end{figure}

In \citetalias{Me}, we argued that the MS heartbeat stars have started to evolve off the MS and that the fraction of the systems that are heartbeat stars rises rapidly with effective temperature. Fig.~\ref{fig:CMD} shows our targets on the \Gaia{} colour-magnitude diagram (CMD), corrected for extinction using the ``Combined19'' dust map with {\tt mwdust} \citep{2016Bovy, 2003Drimmel, 2006Marshall, 2019Green}. The initial impetus behind a search for HBs in APOGEE was to better populate the lower MS with HBs. However, our new sample of HBs does the opposite, extending the population further up the MS. The typical primary is massive ($\sim2$--$8\ M_{\odot}$, Fig.~\ref{fig:Mass}). There are four curves in Fig.~\ref{fig:Line} to help guide the discussion: the Kraft break at $T_{\rm{eff}} = 6550$~K \citep{1967Kraft, 1962Schatzman, 2024Beyer} below which stellar envelopes are convective, a giant branch cut at $M_G > 4 (B_P-R_P-1/2)$ to isolate the MS, a 0.5 Gyr MIST isochrone \citep{2016Dotter, 2016Choi}, and an equal-mass binary isochrone shifted 0.75 mag brighter. 

We binned the stars in colour (this differs from \citetalias{Me} where we used bins parallel to the Kraft break) and computed their median magnitude, as shown in Fig.~\ref{fig:Line}. We also show the corresponding curves from \citetalias{Me} for the HBs detected in the \Gaia{} spectroscopic binary sample. We again find that the binned HBs lie at or above the equal-mass binary isochrone. Yet we know from the discussion of the masses in \S\ref{sec:RV} and the results of \citet{SK2021} and \citet{Sol2025} that many MS HB systems are not twins (Fig.~\ref{fig:Mass}). This adds further support to our hypothesis that many of the HB primaries have started to evolve off the MS.

\begin{figure}
    \begin{tabular}{c}
        \includegraphics[width=\columnwidth]{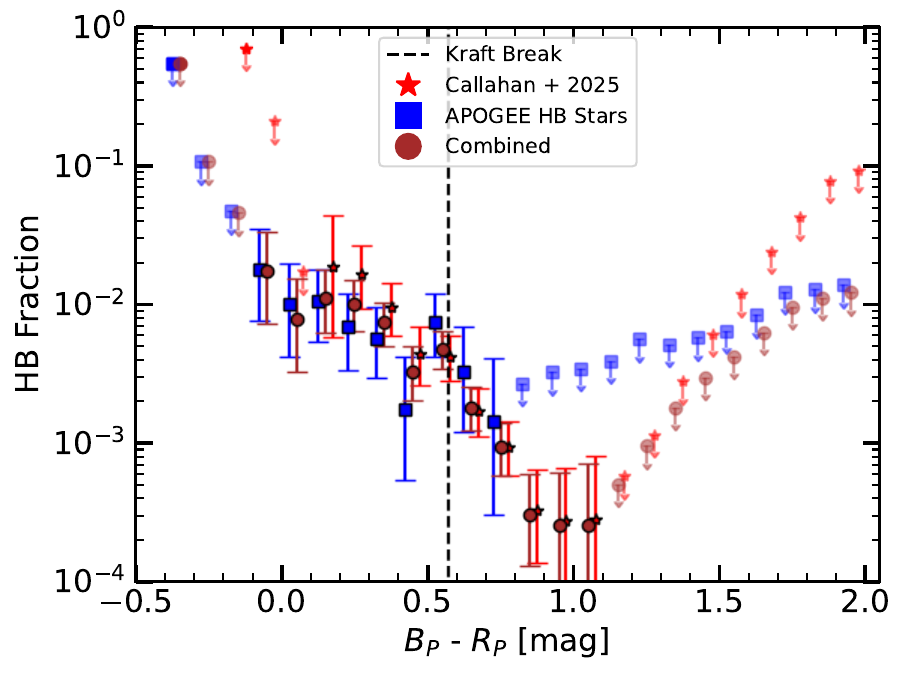} \\       
    \end{tabular}
    \caption{Fraction of HB binaries in 0.1~mag wide colour bins for stars below the giant branch cutoff in Fig.~\ref{fig:Line}. The uncertainties and limits are at 90$\%$ confidence. The Kraft break \citep{1967Kraft, 1962Schatzman, 2024Beyer} divides lower mass stars with convective envelopes from the higher mass stars with radiative envelopes. There are slight (0.025) colour offsets for the APOGEE and \citet{Me} points around the central combined data for visual clarity.}
    \label{fig:Frac}
\end{figure}

Fig.~\ref{fig:Frac} shows the fraction of binaries that are HBs as a function of colour for the APOGEE, \textit{Gaia}, and combined samples. The APOGEE HB fractions decline rapidly with redder colours and are in remarkably good agreement with the \textit{Gaia} SB1 sample \citepalias{Me}. We also show results for the combined sample of APOGEE and \textit{Gaia} HBs. The APOGEE sample confirms the striking result of \citetalias{Me} that the fraction of HBs is highest for the bluest stars and drops rapidly with redder colour down to the Kraft break. We again find that HBs are not rare on the upper MS, where $\sim1\%$ of binaries with $B_P-R_P < 0.5$~mag are HBs.

\begin{table*}
    \renewcommand{\arraystretch}{1.5}
	\centering
	\caption{Properties of the RV orbits fit to eight targets. $N_{\rm{RV}}$ is the number of non-rejected RV epochs. The maximum phase distance, $\min(D_{RV_{0,-1}})$, is the gap in phase between first and last RV measurement in the orbit to measure the percent of the orbit with no phase coverage. $P_{RV}$ is the RV orbital period, $K$ is velocity semi-amplitude, and $\gamma$ is the center-of-mass velocity. The maximum phase distance, $\max(D_\pm)$, is the gap between the measurement and RV quadrature. $f(M)$ is the binary mass function. $M_1$ is the median of the primary mass posterior from StarHorse \citep{2019Anders, 2022Anders}. We use this mass to calculate the secondary mass $M_2$ from the binary mass function (Eq.~\ref{eq:BMF}), and $\sigma_{M_2}$ is the uncertainty on this companion mass. To account for the uncertainty in the primary mass, we also give the 16th--84th percentile range for the $M_1$ posterior from StarHorse and report the secondary masses implied from this range as $M_{2_{16-84}}$.}
    \begin{tabular}{| c  c  c  c  r  r  c  c  c  c  c  c  c |}
        \hline
        TIC & $N_{\rm{RV}}$ & $\max(D_{RV_{0,-1}})$ & $P_{RV}$ & \multicolumn{1}{c}{$K$} & \multicolumn{1}{c}{$\gamma$} & $\max(D_\pm)$ & $f(M)$ & $M_{1}$ & $M_{1_{16-84}}$ & $M_{2}$ & $M_{2_{16-84}}$ & $\sigma_{M_{2}}$\\
           &  &  & (days) & \multicolumn{1}{c}{(km/s)} & \multicolumn{1}{c}{(km/s)} &  & ($M_{\odot}$) & ($M_{\odot}$) & ($M_{\odot}$) & ($M_{\odot}$) & ($M_{\odot}$) & ($M_{\odot}$) \\
        \hline

29453664 & 3 & 0.52 & 3.847 & 64.96 $\pm$ 2.62 & $-$0.61 $\pm$ 2.14 & 0.23 & 0.099 $\pm$ 0.012 & 4.20 & 3.43--4.46 & 1.57 & 1.40--1.63 & $\pm$0.11 \\
54228807 & 3 & 0.47 & 4.037 & 33.69 $\pm$ 2.47 & 38.14 $\pm$ 1.72 & 0.16 & 0.014 $\pm$ 0.003 & 4.12 & 4.09--4.46 & 0.97 & 0.96--1.02 & $\pm$0.09 \\
72193860 & 3 & 0.67 & 5.608 & 26.75 $\pm$ 0.42 & 8.12 $\pm$ 0.33 & 0.17 & 0.011 $\pm$ 0.001 & 2.62 & 2.46--2.75 & 0.68 & 0.67--0.70 & $\pm$0.02 \\
%82355125$^*$ & 4.579 & 1709.69 $\pm$ 1816.00 & 2048.74 $\pm$ 2142.37 & 2178.294 $\pm$ $6941.25$ & 1.75 & 1.68-1.82 & 3759.73 & 3759.58-3759.87 \\
%140810845$^*$ & 3.542 & 80.67 $\pm$ 0.02 & 18.60 $\pm$ 0.01 & 0.171 $\pm$ $1.42\times10^{-4}$ & 2.43 & 2.13-2.68 & 2.54 & 2.38-2.66 \\
%171255871 & 2.089 & 36.86 $\pm$ 2.39 & 12.80 $\pm$ 1.52 & 0.010 $\pm$ $2.02\times10^{-3}$ & 2.06 & 1.91-2.38 & 0.61 & 0.58-0.67 \\
172594580$^*$ & 3 & 0.43 & 10.058 & 64.90 $\pm$ 0.47 & $-$8.21 $\pm$ 0.17 & 0.17 & 0.287 $\pm$ 0.006 & 2.04 & 1.84--2.15 & 4.11 & 3.95--4.20 & $\pm$0.28 \\
184740151 & 3 & 0.41 & 6.871 & 4.87 $\pm$ 0.72 & $-$2.67 $\pm$ 0.39 & 0.15 & 0.00007 $\pm$ 0.00003 & 2.92 & 2.23--3.72 & 0.16 & 0.13--0.18 & $\pm$0.03 \\
229279756 & 4 & 0.55 & 2.938 & 82.60 $\pm$ 0.84 & 34.92 $\pm$ 0.55 & 0.07 & 0.163 $\pm$ 0.005 & 1.94 & 1.76--1.98 & 1.66 & 1.58--1.68 & $\pm$0.06 \\
%281316097$^*$ & 3 & 2.267 & 110.35 $\pm$ 2.35 & 9.65 $\pm$ 0.88 & 0.79 & 0.298 $\pm$ 0.019 & 4.95 & 4.23-5.06 & 8.55 & 8.01-8.63 & $\pm$0.45 \\
%317766125 & 3 & 6.364 & 47.15 $\pm$ 0.60 & $-$8.34 $\pm$ 0.50 & 0.40 & 0.065 $\pm$ 0.002 & 1.92 & 1.87-2.28 & 1.21 & 1.19-1.33 & $\pm$0.03 \\
321952939$^*$ & 3 & 0.56 & 6.654 & 85.13 $\pm$ 0.18 & 4.89 $\pm$ 0.16 & 0.07 & 0.336 $\pm$ 0.002 & 1.83 & 1.77--2.05 & 3.09 & 3.05--3.25 & $\pm$0.33 \\
%348508349$^*$ & 4.146 & 297.03 $\pm$ 0.50 & $-$122.10 $\pm$ 0.23 & 9.703 $\pm$ 0.05 & 6.02 & 5.54-6.75 & 35.68 & 35.04-36.63 \\
390661644 & 4 & 0.53 & 5.068 & 61.75 $\pm$ 2.05 & 9.12 $\pm$ 1.46 & 0.13 & 0.117 $\pm$ 0.012 & 4.94 & 4.65--5.32 & 2.46 & 2.38--2.57 & $\pm$0.12 \\
%444446485$^*$ & 3.408 & 89.17 $\pm$ 0.32 & $-$28.63 $\pm$ 0.06 & 0.216 $\pm$ $2.32\times10^{-3}$ & 1.78 & 1.51-1.96 & 2.65 & 2.46-2.76 \\

%$7.46\times10^{-5}$ $\pm$ $3.29\times10^{-5}$

\hline
\end{tabular}
	\label{tab:table_rv}
    \begin{tablenotes}
      \small
      \item * Targets where the secondary mass is larger than the primary mass (Fig.~\ref{fig:Mass})
    \end{tablenotes}
\end{table*}

Upcoming data releases from large spectroscopic surveys like SDSS-V Milky Way Mapper and \Gaia{} DR4 will substantially expand the samples available for detecting and characterizing rare binary populations like HBs. With epoch RV measurements from \Gaia{} DR4, for example, there will be many more double-lined spectroscopic binary solutions that can be used to measure precise masses and radii of HBs. Additionally, \textit{TESS} Cycle 8 includes longer 54-day sectors that could be used to detect longer-period HBs with $P>13$~days. Alternatively, heavily binned light curves from ground-based surveys like the All-Sky Automated Survey for Supernovae \citep[ASAS-SN, ][]{2014Shappee, 2017Kochanek} could be used to verify the periods and detect the very weak signals of Kepler-like HBs \citep{2025Hon}.

%The results of our APOGEE analysis act to back up the prior results continuing the support the idea that HBs are not just binaries, but binaries in which the primary has started to evolve. It remains clear from \citetalias{Me} that there is considerable physics in characterizing and understanding the distribution of HBs as a function of orbital and stellar properties. There are, however, a large number of variables (mass, evolutionary state, orbit, etc.), which means that additional samples completing the MS will be needed to characterize the physical parameter space of HBs~fully.

%\input ANC/Full_table/APOGEE_Table.tex

\section*{Acknowledgments}
JC thanks Krzysztof Stanek for useful discussions in the course of our research. CSK is supported by NSF grants AST-2307385 and 2407206. ASAS-SN is funded by Gordon and Betty Moore Foundation grants GBMF5490 and GBMF10501 and the Alfred P. Sloan Foundation grant G-2021-14192.

Support for this work was provided by NASA through the NASA Hubble Fellowship grant HST-HF2-51588.001-A awarded by the Space Telescope Science Institute, which is operated by the Association of Universities for Research in Astronomy, Inc., for NASA, under contract
NAS5-26555.

Funding for the Sloan Digital Sky Survey IV has been provided by the Alfred P. Sloan Foundation, the U.S. Department of Energy Office of Science, and the Participating Institutions. 

SDSS-IV acknowledges support and resources from the Center for High Performance Computing  at the University of Utah. The SDSS website is \url{www.sdss4.org}.

SDSS-IV is managed by the Astrophysical Research Consortium for the Participating Institutions of the SDSS Collaboration including the Brazilian Participation Group, the Carnegie Institution for Science, Carnegie Mellon University, Center for Astrophysics | Harvard \& Smithsonian, the Chilean Participation Group, the French Participation Group, Instituto de Astrof\'isica de Canarias, The Johns Hopkins University, Kavli Institute for the Physics and Mathematics of the Universe (IPMU) / University of Tokyo, the Korean Participation Group, Lawrence Berkeley National Laboratory, Leibniz Institut f\"ur Astrophysik Potsdam (AIP),  Max-Planck-Institut f\"ur Astronomie (MPIA Heidelberg), Max-Planck-Institut f\"ur Astrophysik (MPA Garching), Max-Planck-Institut f\"ur Extraterrestrische Physik (MPE), National Astronomical Observatories of China, New Mexico State University, New York University, University of Notre Dame, Observat\'ario Nacional / MCTI, The Ohio State University, Pennsylvania State University, Shanghai Astronomical Observatory, United Kingdom Participation Group, Universidad Nacional Aut\'onoma de M\'exico, University of Arizona, University of Colorado Boulder, University of Oxford, University of Portsmouth, University of Utah, University of Virginia, University of Washington, University of Wisconsin, Vanderbilt University, and Yale University.

This paper also includes data collected by the TESS mission. Funding for the TESS mission is provided by the NASA Science Mission Directorate.

\clearpage
\bibliographystyle{mnras}
\bibliography{hbs} % if your bibtex file is called example.bib

\end{document}